\documentclass[longauth]{aa} 

\usepackage{graphicx}
\usepackage{txfonts}
\usepackage{hyperref}
\usepackage{booktabs}
\usepackage{academicons}
\usepackage{xcolor}
\usepackage{fontawesome}
\usepackage{xcolor}
\usepackage[normalem]{ulem}

\begin{document}

\title{Refined parameters of the HD\,22946 planetary system\\ and the true orbital period of planet d\thanks{This article uses data from CHEOPS programmes CH\_PR110048 and CH\_PR100031. Photometry and radial velocity data of HD\,22946 are available at the CDS via anonymous ftp to \url{cdsarc.u-strasbg.fr} (130.79.128.5) or via \url{http://cdsarc.u-strasbg.fr/viz-bin/qcat?J/A+A/}.}}

\author{
Z.~Garai\inst{\ref{inst:001},\ref{inst:002},\ref{inst:003}},
H.~P.~Osborn\inst{\ref{inst:004},\ref{inst:005}},
D.~Gandolfi\inst{\ref{inst:006}},
A.~Brandeker\inst{\ref{inst:007}},
S.~G.~Sousa\inst{\ref{inst:008}},
M.~Lendl\inst{\ref{inst:012}},
A.~Bekkelien\inst{\ref{inst:012}},
C.~Broeg\inst{\ref{inst:004},\ref{inst:025}},
A.~Collier~Cameron\inst{\ref{inst:013}},
J.~A.~Egger\inst{\ref{inst:004}},
M.~J.~Hooton\inst{\ref{inst:004},\ref{inst:009}},
Y.~Alibert\inst{\ref{inst:004}},
L.~Delrez\inst{\ref{inst:010},\ref{inst:028}},
L.~Fossati\inst{\ref{inst:011}},
S.~Salmon\inst{\ref{inst:012}},
T.~G.~Wilson\inst{\ref{inst:013}},
A.~Bonfanti\inst{\ref{inst:014}},
A.~Tuson\inst{\ref{inst:009}},
S.~Ulmer-Moll\inst{\ref{inst:004},\ref{inst:012}},
L.~M.~Serrano\inst{\ref{inst:006}},
L.~Borsato\inst{\ref{inst:015}},
R.~Alonso\inst{\ref{inst:019},\ref{inst:029}},
G.~Anglada\inst{\ref{inst:020},\ref{inst:030}},
J.~Asquier\inst{\ref{inst:018}},
D.~Barrado~y~Navascues\inst{\ref{inst:021}},
S.~C.~C.~Barros\inst{\ref{inst:008},\ref{inst:024}},
T.~Bárczy\inst{\ref{inst:037}},
W.~Baumjohann\inst{\ref{inst:011}},
M.~Beck\inst{\ref{inst:012}},
T.~Beck\inst{\ref{inst:004}},
W.~Benz\inst{\ref{inst:004},\ref{inst:025}},
N. Billot\inst{\ref{inst:012}},
F.~Biondi\inst{\ref{inst:015},\ref{inst:033}},
X.~Bonfils\inst{\ref{inst:022}},
M.~Buder\inst{\ref{inst:017}},
J.~Cabrera\inst{\ref{inst:017}},
V.~Cessa\inst{\ref{inst:004}},
S.~Charnoz\inst{\ref{inst:023}},
Sz.~Csizmadia\inst{\ref{inst:017}},
P.~E.~Cubillos\inst{\ref{inst:014},\ref{inst:044}},
M.~B.~Davies\inst{\ref{inst:040}},
M.~Deleuil\inst{\ref{inst:031}},
O.~D.~S.~Demangeon\inst{\ref{inst:022},\ref{inst:024}},
B.-O.~Demory\inst{\ref{inst:025}},
D.~Ehrenreich\inst{\ref{inst:012}},
A.~Erikson\inst{\ref{inst:017}},
V.~Van~Eylen\inst{\ref{inst:045}},
A.~Fortier\inst{\ref{inst:004},\ref{inst:025}},
M.~Fridlund\inst{\ref{inst:026},\ref{inst:027}},
M.~Gillon\inst{\ref{inst:010}},
V.~Van~Grootel\inst{\ref{inst:028}},
M.~Güdel\inst{\ref{inst:016}},
M.~N.~Günther\inst{\ref{inst:018}},
S.~Hoyer\inst{\ref{inst:031}},
K.~G.~Isaak\inst{\ref{inst:018}},
L.~L.~Kiss\inst{\ref{inst:038}},
M.~H.~Kristiansen\inst{\ref{inst:046}},
J.~Laskar\inst{\ref{inst:032}},
A.~Lecavelier~des~Etangs\inst{\ref{inst:039}},
C.~Lovis\inst{\ref{inst:012}},
A.~Luntzer\inst{\ref{inst:016}},
D.~Magrin\inst{\ref{inst:015}},
P.~F.~L.~Maxted\inst{\ref{inst:036}},
C.~Mordasini\inst{\ref{inst:004}},
V.~Nascimbeni\inst{\ref{inst:015}},
G.~Olofsson\inst{\ref{inst:007}},
R.~Ottensamer\inst{\ref{inst:016}},
I.~Pagano\inst{\ref{inst:035}},
E.~Pallé\inst{\ref{inst:019},\ref{inst:029}},
G.~Peter\inst{\ref{inst:017}},
G.~Piotto\inst{\ref{inst:015},\ref{inst:034}},
D.~Pollacco\inst{\ref{inst:041}},
D.~Queloz\inst{\ref{inst:009},\ref{inst:012}},
R.~Ragazzoni\inst{\ref{inst:015},\ref{inst:034}},
N.~Rando\inst{\ref{inst:018}},
H.~Rauer\inst{\ref{inst:017},\ref{inst:042}},
I.~Ribas\inst{\ref{inst:020},\ref{inst:030}},
N.~C.~Santos\inst{\ref{inst:008},\ref{inst:024}},
G.~Scandariato\inst{\ref{inst:035}},
D.~Ségransan\inst{\ref{inst:012}},
A.~E.~Simon\inst{\ref{inst:004}},
A.~M.~S.~Smith\inst{\ref{inst:017}},
M.~Steller\inst{\ref{inst:011}},
Gy.~M.~Szabó\inst{\ref{inst:001},\ref{inst:002}},
N.~Thomas\inst{\ref{inst:004}},
S.~Udry\inst{\ref{inst:012}},
J.~Venturini\inst{\ref{inst:012}},
\and
N.~Walton\inst{\ref{inst:043}}
}

\institute{
\label{inst:001} MTA-ELTE Exoplanet Research Group, 9700 Szombathely, Szent Imre h. u. 112, Hungary, \email{zgarai@gothard.hu} \and
\label{inst:002} ELTE Gothard Astrophysical Observatory, 9700 Szombathely, Szent Imre h. u. 112, Hungary \and
\label{inst:003} Astronomical Institute, Slovak Academy of Sciences, 05960 Tatransk\'a Lomnica, Slovakia \and
\label{inst:004} Physikalisches Institut, University of Bern, Gesellsschaftstrasse 6, 3012 Bern, Switzerland \and
\label{inst:005} Department of Physics and Kavli Institute for Astrophysics and Space Research, Massachusetts Institute of Technology, Cambridge, MA 02139, USA \and
\label{inst:006} Dipartimento di Fisica, Universita degli Studi di Torino, via Pietro Giuria 1, I-10125, Torino, Italy \and
\label{inst:007} Department of Astronomy, Stockholm University, AlbaNova University Center, 10691 Stockholm, Sweden \and
\label{inst:008} Instituto de Astrofisica e Ciencias do Espaco, Universidade do Porto, CAUP, Rua das Estrelas, 4150-762 Porto, Portugal \and
\label{inst:009} Astrophysics Group, Cavendish Laboratory, University of Cambridge, J.J. Thomson Avenue, Cambridge CB3 0HE, UK \and
\label{inst:010} Astrobiology Research Unit, Université de Liège, Allée du 6 Août 19C, B-4000 Liège, Belgium \and
\label{inst:011} Space Research Institute, Austrian Academy of Sciences, Schmiedlstrasse 6, A-8042 Graz, Austria \and
\label{inst:012} Observatoire Astronomique de l'Université de Genève, Chemin Pegasi 51, Versoix, Switzerland \and
\label{inst:013} Centre for Exoplanet Science, SUPA School of Physics and Astronomy, University of St Andrews, North Haugh, St Andrews KY16 9SS, UK \and
\label{inst:014} Space Research Institute, Austrian Academy of Sciences, Schmiedlstrasse 6, 8042 Graz, Austria \and
\label{inst:015} INAF, Osservatorio Astronomico di Padova, Vicolo dell'Osservatorio 5, 35122 Padova, Italy \and
\label{inst:016} Department of Astrophysics, University of Vienna, Tuerkenschanzstrasse 17, 1180 Vienna, Austria \and
\label{inst:017} Institute of Planetary Research, German Aerospace Center (DLR), Rutherfordstrasse 2, 12489 Berlin, Germany \and
\label{inst:018} European Space Agency (ESA), European Space Research and Technology Centre (ESTEC), Keplerlaan 1, 2201 AZ Noordwijk, The Netherlands \and
\label{inst:019} Instituto de Astrofisica de Canarias, 38200 La Laguna, Tenerife, Spain \and
\label{inst:020} Institut d'Estudis Espacials de Catalunya (IEEC), 08034 Barcelona, Spain \and
\label{inst:021} Depto. de Astrofisica, Centro de Astrobiologia (CSIC-INTA), ESAC campus, 28692 Villanueva de la Cañada (Madrid), Spain \and
\label{inst:022} Université Grenoble Alpes, CNRS, IPAG, 38000 Grenoble, France \and
\label{inst:023} Université de Paris, Institut de physique du globe de Paris, CNRS, F-75005 Paris, France \and
\label{inst:024} Departamento de Fisica e Astronomia, Faculdade de Ciencias, Universidade do Porto, Rua do Campo Alegre, 4169-007 Porto, Portugal \and
\label{inst:025} Center for Space and Habitability, University of Bern, Gesellschaftsstrasse 6, 3012, Bern, Switzerland \and
\label{inst:026} Leiden Observatory, University of Leiden, PO Box 9513, 2300 RA Leiden, The Netherlands \and
\label{inst:027} Department of Space, Earth and Environment, Chalmers University of Technology, Onsala Space Observatory, 43992 Onsala, Sweden \and
\label{inst:028} Space sciences, Technologies and Astrophysics Research (STAR) Institute, Universit\'e de Li\`ege, 19C All\'ee du 6 Ao\^ut, B-4000 Li\`ege, Belgium \and
\label{inst:029} Departamento de Astrofisica, Universidad de La Laguna, 38206 La Laguna, Tenerife, Spain \and
\label{inst:030} Institut de Ciencies de l'Espai (ICE, CSIC), Campus UAB, Can Magrans s/n, 08193 Bellaterra, Spain \and
\label{inst:031} Aix Marseille Univ, CNRS, CNES, LAM, 38 rue Frédéric Joliot-Curie, 13388 Marseille, France \and
\label{inst:032} IMCCE, UMR8028 CNRS, Observatoire de Paris, PSL Univ., Sorbonne Univ., 77 av. Denfert-Rochereau, 75014 Paris, France \and
\label{inst:033} Max Planck Institute for Extraterrestrial Physics, Giessenbachstrasse 1, 85748 Garching bei München, Germany \and
\label{inst:034} Dipartimento di Fisica e Astronomia "Galileo Galilei", Universita degli Studi di Padova, Vicolo dell'Osservatorio 3, 35122 Padova, Italy \and
\label{inst:035} INAF, Osservatorio Astrofisico di Catania, Via S. Sofia 78, 95123 Catania, Italy \and
\label{inst:036} Astrophysics Group, Keele University, Staffordshire, ST5 5BG, United Kingdom \and
\label{inst:037} Admatis, 5. Kandó Kálmán Street, 3534 Miskolc, Hungary \and
\label{inst:038} Konkoly Observatory, Research Centre for Astronomy and Earth Sciences, 1121 Budapest, Konkoly Thege Miklós út 15-17, Hungary \and
\label{inst:039} Institut d'astrophysique de Paris, UMR7095 CNRS, Université Pierre \& Marie Curie, 98bis blvd. Arago, 75014 Paris, France \and
\label{inst:040} Centre for Mathematical Sciences, Lund University, Box 118, 22100 Lund, Sweden \and
\label{inst:041} Department of Physics, University of Warwick, Gibbet Hill Road, Coventry CV4 7AL, United Kingdom \and
\label{inst:042} Center for Astronomy and Astrophysics, Technical University Berlin, Hardenberstrasse 36, 10623 Berlin, Germany \and
\label{inst:043} Institute of Astronomy, University of Cambridge, Madingley Road, Cambridge, CB3 0HA, United Kingdom \and
\label{inst:044} INAF, Osservatorio Astrofisico di Torino, Via Osservatorio, 20, 10025 Pino Torinese TO, Italy \and
\label{inst:045} Department of Space and Climate Physics, Mullard Space Science Laboratory, Holmbury St Mary RH5 6NT, UK \and
\label{inst:046} Brorfelde Observatory, Observator Gyldenkernes Vej 7, DK-4340 Tølløse, Denmark 
}

\date{Received September 15, 1996; accepted March 16, 1997}

 
\abstract
{Multi-planet systems are important sources of information regarding the evolution of planets. However, the long-period planets  in these systems often escape detection. These objects in particular may retain more of their primordial characteristics compared to close-in counterparts because of their increased distance from the host star. HD\,22946 is a bright ($G=8.13$ mag) late F-type star around which three transiting planets were identified via Transiting Exoplanet Survey Satellite (TESS) photometry, but the true orbital period of the outermost planet d was unknown until now.}
{We aim to use the Characterising Exoplanet Satellite (CHEOPS) space telescope to uncover the true orbital period of HD\,22946d and to refine the orbital and planetary properties of the system, especially the radii of the planets.}
{We used the available TESS photometry of HD\,22946 and observed several transits of the planets b, c, and d using CHEOPS. We identified two transits of planet d in the TESS photometry, calculated the most probable period aliases based on these data, and then scheduled CHEOPS observations. The photometric data were supplemented with ESPRESSO (Echelle SPectrograph for Rocky Exoplanets and Stable Spectroscopic Observations) radial velocity data. Finally, a combined model was fitted to the entire dataset in order to obtain final planetary and system parameters.}
{Based on the combined TESS and CHEOPS observations, we successfully determined the true orbital period of the planet d to be $47.42489 \pm 0.00011$ d, and derived precise radii of the planets in the system, namely $1.362 \pm 0.040~\mathrm{R_\oplus}$, $2.328 \pm 0.039~\mathrm{R_\oplus}$, and $2.607 \pm 0.060~\mathrm{R_\oplus}$ for planets b, c, and d, respectively. Due to the low number of radial velocities, we were only able to determine 3$\sigma$ upper limits for these respective planet masses, which are $13.71~\mathrm{M}_\oplus$, $9.72~\mathrm{M}_\oplus$, and $26.57~\mathrm{M}_\oplus$. We estimated that another 48 ESPRESSO radial velocities are needed to measure the predicted masses of all planets in HD 22946. We also derived stellar parameters for the host star.}
{Planet c around HD\,22946 appears to be a promising target for future atmospheric characterisation via transmission spectroscopy. We can also conclude that planet d, as a warm sub-Neptune, is very interesting because there are only a few similar confirmed exoplanets to date. Such objects are worth investigating in the near future, for example in terms of their composition and internal structure.}


\titlerunning{Refined parameters of the HD\,22946 system}
\authorrunning{Z. Garai et al.}
\maketitle

\section{Introduction}
\label{intro}

Multi-planet systems are important from many viewpoints. Not only are they susceptible of relatively straightforward confirmation as bona fide planets \citep{lissauer2012almost}, they also allow intra-planetary comparisons to be made for planets which formed under the same conditions; see for example \citet{weiss2018california}. The majority of the known multi-planet systems were found by space-based exoplanet transit surveys. This is because, while giant hot-Jupiters are relatively easy to observe with ground-based photometry, the detection of smaller planets, for example, Earths, super-Earths, and sub-Neptunes, which are typically found in multi-planet systems, requires the precise photometry of space-based observatories such as TESS \citep{Ricker1}.

Mutual gravitational interactions in some multi-planet systems can provide constraints on the planet masses through transit time variations (TTVs); see for example \citet{Nesvorny1}. Alternatively, radial velocity (RV) observations are needed to put constraints on the masses of planets \citep{Mayor1}. Even where masses cannot be determined, mass upper limits can provide proof that the studied objects are of planetary origin; see for example \citet{Hord1}, \citet{Wilson1}, or \citet{Stefansson1}. Mass determination can then help constrain the internal structure of the planet bodies, and break degeneracies in atmospheric characterisation follow-up studies. If precise planet radii are also determined from transit photometry, this allows  the planet internal density to be calculated and the planetary composition to be estimated; see for example \citet{Delrez1} and \citet{Lacedelli1, Lacedelli2}. Precise planetary parameters also allow the planets to be put in the context of population trends, such as the radius \citep{Fulton1, vanEylen1, Martinez1, Ho1} and density \citep{luque2022density} valleys. 
 
Long-period planets in multiple-planet systems often escape detection, especially when their orbital periods are longer than the typical observing duration of photometric surveys (e.g. $\sim 27$ d for TESS). However, detecting such planets is also important. For example, the increased distance from their host stars means that, when compared with close-in planets, they may retain more of their primordial characteristics, such as unevaporated atmospheres \citep{Owen1} or circumplanetary material \citep{Dobos1}. Due to the limited observing duration of the TESS primary mission, which observed the majority of the near-ecliptic sectors for only 27 days, planets on long periods produce only single transits. However, thanks to its extended mission, TESS re-observed the same fields two years later, and in many cases was able to re-detect a second transit; see for example \citet{Osborn1}. These `duotransit' cases require follow-up in order to uncover the true orbital period due to the gap, which causes a set of aliases, $P \in (t_\mathrm{tr,2} - t_\mathrm{tr,1})/(1,2,3,\dots,N_\mathrm{max})$, where $t_\mathrm{tr,1}$ and $t_\mathrm{tr,2}$ are the first and the second observed mid-transit times, respectively. The longest possible period is the temporal distance between the two mid-transit times, $P_\mathrm{max} = (t_\mathrm{tr,2} - t_\mathrm{tr,1})$, and the shortest possible period is bounded by the non-detection of subsequent transits. 

In addition to ground-based telescopes, the CHEOPS space observatory \citep{Benz1} can be used to follow-up duotransit targets and to determine their true orbital periods and other characteristics. For example, the periods of two young sub-Neptunes orbiting BD+40 2790 (TOI-2076, TIC-27491137) were found using a combination of CHEOPS and ground-based photometric follow-up observations \citep{Osborn1}. Furthermore, these combined observations uncovered the TTVs of two planets, and also improved the radius precision of all planets in the system. CHEOPS observations also recovered orbital periods of duotransits in HIP 9618 \citep{Osborn_sub}, TOI-5678 \citep{Ulmer-Moll_sub}, and HD 15906 \citep{Tuson_sub} systems. In the present study, we investigated the HD\,22946 system with a similar aim. HD\,22946 (TOI-411, TIC-100990000) is a bright ($G = 8.13$ mag) late F-type star with three transiting planets. The planetary system was discovered and validated only recently by \citet{Cacci2}; hereafter C22. The authors presented several parameters of the system, including the radii and mass limits of the planets. They found that planet b is a super-Earth with a radius of $1.72 \pm 0.10~\mathrm{R}_\oplus$, while planets c and d are sub-Neptunes with radii of $2.74 \pm 0.14~\mathrm{R}_\oplus$ and $3.23 \pm 0.19~\mathrm{R}_\oplus$, respectively. The $3\sigma$ upper mass limits of planets  b, c, and d were determined ---based on ESPRESSO spectroscopic observations (see Sect. \ref{espdata})--- to be $11~\mathrm{M}_\oplus$, $14.5~\mathrm{M}_\oplus$, and $24.5~\mathrm{M}_\oplus$, respectively. As TESS recorded several transits during observations in sector numbers 3, 4, 30, and 31, the discoverers easily derived the orbital periods of the two inner planets, b and c, which are about 4.040 d and 9.573 d, respectively. The orbital period of planet d was not found by C22. The authors determined its presence through a single transit found in sector number 4 and obtained its parameters from this single transit event. Its depth and the host brightness make planet d easily detectable with CHEOPS, and therefore HD\,22946 was observed several times with this instrument within the Guaranteed Time Observations (GTO) programmes CH\_PR110048 and CH\_PR100031, with the main scientific goals being to uncover the true orbital period of planet d and to refine the parameters of the HD\,22946 system based on CHEOPS and TESS observations via joint analysis of the photometric data, supplemented with ESPRESSO spectroscopic observations of HD\,22946.  

The present paper is organised as follows. In Sect. \ref{secobs}, we provide a brief description of observations and data reduction. In Sect. \ref{dataan}, we present the details of our data analysis and our first results, including stellar parameters, period aliases of HD\,22946d from the TESS data, and a search for TTVs. Our final results based on the combined TESS, CHEOPS, and RV model are described and discussed in Sect. \ref{res}. We summarise our findings in Sect. \ref{conc}.               

\begin{table}
\centering
\caption{Log of TESS photometric observations of HD\,22946.}
\label{TESSobslog}
\begin{tabular}{ccc}
\hline
\hline
Time interval of observation    & Sector No.    & Transits\\
\hline
\hline
\multicolumn{3}{c}{HD\,22946b}\\
2018-09-20 -- 2018-10-18        & 03                    & 5\\
2018-10-18 -- 2018-11-15        & 04                    & 6\\
2020-09-22 -- 2020-10-21        & 30                    & 6\\
2020-10-21 -- 2020-11-19        & 31                    & 6\\
\hline
\multicolumn{3}{c}{HD\,22946c}\\
2018-09-20 -- 2018-10-18        & 03                    & 2\\
2018-10-18 -- 2018-11-15        & 04                    & 2\\
2020-09-22 -- 2020-10-21        & 30                    & 2\\
2020-10-21 -- 2020-11-19        & 31                    & 2\\
\hline
\multicolumn{3}{c}{HD\,22946d}\\
2018-10-18 -- 2018-11-15        & 04                    & 1\\
2020-09-22 -- 2020-10-21        & 30                    & 1\\
\hline
\hline
\end{tabular}
\end{table}

\section{Observations and data reduction}
\label{secobs}

\subsection{TESS data}
\label{tessdata}

HD\,22946 was observed during four TESS sectors: numbers 3, 4, 30, and 31 (see Table \ref{TESSobslog}). The time gap between the two observing seasons is almost two years. These TESS data were downloaded from the Mikulski Archive for Space Telescopes\footnote{See \url{https://mast.stsci.edu/portal/Mashup/Clients/Mast/Portal.html}.} in the form of Presearch Data Conditioning Simple Aperture Photometry (PDCSAP) flux. These data, containing 61~987 data points, were obtained from two-minute integrations and were initially smoothed by the PDCSAP pipeline. This light curve is subjected to more treatment than the simple aperture photometry (SAP) light curve, and is specifically intended for detecting planets. The pipeline attempts to remove systematic artifacts while keeping planetary transits intact. The average uncertainty of the PDCSAP data points is 310 ppm. 

During these TESS observing runs, 23 transits of planet b were recorded, and the transit of planet c was observed eight times in total (see more details in Table \ref{TESSobslog}). As in C22, we also initially recognised a transit-like feature in the sector number 4 data at $t_\mathrm{tr,1} = 2~458~425.1657$ $\mathrm{BJD}_\mathrm{TDB}$ through visual inspection of the light curve. Given 65\%--80\% of single transits from the TESS primary mission will re-transit in the extended mission sectors \citep[see][]{Cooke2019, Cooke2021}, we subsequently visually inspected the light curve once the TESS year 3 data were available and found a second dip at $t_\mathrm{tr,2} = 2~459~136.5357$ $\mathrm{BJD}_\mathrm{TDB}$ in the sector number 30 data with near-identical depth and duration. Given the high prior probability of finding a second transit, the close match in transit shape between events, and the high quality of the data (i.e. minimal systematic noise elsewhere in the light curve), we concluded that this signal is a bona fide transit event and that the transits in sector numbers 4 and 30 are very likely caused by the same object, that is, by planet d.

Outliers were cleaned using a $3\sigma$ clipping, where $\sigma$ is the standard deviation of the light curve. With this clipping procedure, we discarded 300 data points out of 61~987, which is $\sim 0.5\%$ of the TESS data. Subsequently, we visually inspected the dataset in order to check the effect of the outlier removal, which we found to be reasonable. As TESS uses as time stamps Barycentric TESS Julian Date (i.e. $\mathrm{BJD}_\mathrm{TDB} - 2~457~000.0$), during the next step we converted all TESS time stamps to $\mathrm{BJD}_\mathrm{TDB}$.      

\subsection{CHEOPS data}
\label{cheopsdata}

HD\,22946 was observed five times with the CHEOPS space telescope. This is the first European space mission dedicated primarily to the study of known exoplanets. It consists of a telescope with a mirror of  32 cm  in diameter based on a Ritchey-Chr\'{e}tien design. The photometric detector is a single-CCD camera covering the wavelength range from 330 to 1100 nm with a field of view of 0.32 deg$^2$. The payload design and operation have been optimised to achieve ultra-high photometric stability, achieving a photometric precision of 20 ppm on observations of a G5-type star in 6 hours, and 85 ppm observations of a K5-type star in 3 hours \citep{Benz1}. The CHEOPS observations were scheduled based on the existing TESS observations of planets b and c, and mainly based on the observed transit times of planet d (see Sect. \ref{tessdata}). The marginal probability for each period alias of planet d was calculated using the \texttt{MonoTools} package (see Sect. \ref{dataanper}). We were not able to observe all the highest-probability aliases, because some were not visible during the two-week period of visibility. Within the program number CH\_PR110048, we therefore planned to observe the three highest-probability aliases of planet d with CHEOPS, but due to observability constraints and conflicts with other observations, only two visits\footnote{A visit is a sequence of successive CHEOPS orbits devoted to observing a given target.} of planet d aliases were scheduled. Its true orbital period was confirmed during the second observation. The remaining three visits were scheduled in the framework of the program number CH\_PR100031. Based on these CHEOPS observations, three transits of planet b were recorded during  visits 1, 3, and 5, the transit of planet c was observed twice during visits 2 and 4, and a single transit of planet d (in multiple transit feature with planet c) was detected during the CHEOPS visit 4. Further details about these observations can be found in Table \ref{cheopsobslog}. 

\begin{table*}
\centering
\caption{Log of CHEOPS photometric observations of HD\,22946.}
\label{cheopsobslog}
\begin{tabular}{ccccccc}
\hline
\hline
Visit   & Start date                    & End date                      & File                                                    & CHEOPS                & Integration             & Number \\
No.     & [UTC]                         & [UTC]                                 & key                                                     & product               & time [s]                & of frames\\   
\hline
\hline
1               & 2021-10-17 03:22              & 2021-10-17 14:40              & {\tt{CH\_PR100031\_TG021201}}           & Subarray              & $2 \times 20.0$   & 629\\
1               & 2021-10-17 03:22              & 2021-10-17 14:40              & {\tt{CH\_PR100031\_TG021201}}           & Imagettes             & 20.0                  & 1258\\
2               & 2021-10-18 08:14              & 2021-10-18 19:04              & {\tt{CH\_PR100031\_TG021101}}           & Subarray              & $2 \times 20.0$   & 637\\
2               & 2021-10-18 08:14              & 2021-10-18 19:04              & {\tt{CH\_PR100031\_TG021101}}           & Imagettes             & 20.0                  & 1274\\
3               & 2021-10-25 07:08              & 2021-10-25 19:49              & {\tt{CH\_PR110048\_TG010001}}           & Subarray              & $2 \times 20.4$   & 708\\
3               & 2021-10-25 07:08              & 2021-10-25 19:49              & {\tt{CH\_PR110048\_TG010001}}           & Imagettes             & 20.4                  & 1416\\
4               & 2021-10-28 02:12              & 2021-10-28 13:50              & {\tt{CH\_PR110048\_TG010101}}           & Subarray              & $2 \times 20.4$   & 666\\ 
4               & 2021-10-28 02:12              & 2021-10-28 13:50              & {\tt{CH\_PR110048\_TG010101}}           & Imagettes             & 20.4                  & 1332\\
5               & 2021-10-29 08:48              & 2021-10-29 18:14              & {\tt{CH\_PR100031\_TG021202}}           & Subarray              & $2 \times 20.0$   & 555\\
5               & 2021-10-29 08:48              & 2021-10-29 18:14              & {\tt{CH\_PR100031\_TG021202}}           & Imagettes             & 20.0                  & 1110\\
\hline
\hline
\end{tabular}
\tablefoot{Table shows the time interval of individual observations, the file key, which supports fast identification of the observations in the CHEOPS archive, type of the photometric product (for more details see Sect. \ref{cheopsdata}), the applied integration time, co-added exposures at the subarray type CHEOPS product, and the number of obtained frames.}
\end{table*}

From the CHEOPS detector, which has $1024 \times 1024$ pixels, a $200 \times 200$ pixels subarray is extracted around the target point-spread function (PSF), which is used to compute the photometry. This type of photometry product was processed by the CHEOPS Data Reduction Pipeline ({DRP}) version 13.1.0 \citep{Hoyer1}. It performs several image corrections, including bias-, dark-, and flat-corrections, contamination estimation, and background-star correction. The \texttt{DRP} pipeline produces four different light-curve types for each visit, but we initially analysed only the decontaminated `\texttt{OPTIMAL}' type, where the aperture radius is automatically set based on the signal-to-noise ratio ($S/N$). In addition to the subarrays, there are imagettes available for each exposure. The imagettes are frames of 30 pixels in radius centred on the target, which do not need to be co-added before download owing to their smaller size. We used a tool specifically developed for photometric extraction of imagettes using point-spread function photometry, called \texttt{PIPE}\footnote{See \url{https://github.com/alphapsa/PIPE}.}; see for example \citet{Szabo1, Szabo2}. The \texttt{PIPE} photometry has a $S/N$ comparable to that of \texttt{DRP} photometry, but has the advantage of shorter cadence, and therefore we decided to use this CHEOPS product in this work. The average uncertainty of the \texttt{PIPE} data points is 160 ppm.   

The \texttt{PIPE} CHEOPS observations were processed using the dedicated data decorrelation and transit analysis software called \texttt{pycheops}\footnote{See \url{https://github.com/pmaxted/pycheops}.} \citep{Maxted1}. This package includes downloading, visualising, and decorrelating CHEOPS data, fitting transits and eclipses of exoplanets, and calculating light-curve noise. We first cleaned the light curves from outlier data points using the {\tt{pycheops}} built-in function {\tt{clip\_outliers}}, which removes outliers from a dataset by calculating the mean absolute deviation ($MAD$) from the light curve following median smoothing, and rejects data greater than the smoothed dataset plus the $MAD$ multiplied by a clipping factor. The clipping factor equal to five was reasonable in our cases, which we checked visually. With this clipping procedure, we discarded 30 data points out of 3195, which is $\sim 0.9\%$ of the CHEOPS data. The next step was the extraction of the detrending parameters. During this procedure, the software gives a list of the parameters necessary for the detrending. The most important decorrelation is
subtraction of the roll-angle effect. In order to keep the cold plate radiators facing away from the Earth, the spacecraft rolls during its orbit. This means that the field of view rotates around the pointing direction. The target star remains stationary within typically 1 pixel, but the rotation of the field of view produces a variation of its flux from the nearby sources in phase with the roll angle of the spacecraft \citep{bonfanti21}. The extracted detrending parameters were co-fitted with the transit model (see Sect. \ref{dataancheopstess}).

\subsection{ESPRESSO/VLT data}
\label{espdata}

We acquired 14 high-resolution spectra of the host star HD\,22946 using the ESPRESSO spectrograph \citep{Pepe2} mounted at the 8.2\,m Very Large Telescope (VLT) at Paranal Observatory (Chile). The observations were carried out between 10 February 2019 and 17 March 2019 under the observing program number 0102.C-0456 (PI: V. Van Eylen) and within the KESPRINT\footnote{See \url{https://kesprint.science/}.} project. We used the high-resolution (HR) mode of the spectrograph, which provides a resolving power of $R\,\approx\,134~000$. We set the exposure time to 600\,s, leading to a $S/N$ per pixel at 650\,nm ranging between 120 and 243. Daytime ThAr spectra and simultaneous Fabry-Perot exposures were taken to determine the wavelength solution and correct for possible nightly instrumental drifts, respectively. We reduced the ESPRESSO spectra using the dedicated data-reduction software and extracted the RVs by cross-correlating the \'echelle spectra with a G2 numerical mask. We list the ESPRESSO RV measurements in Table \ref{espdatatable}. The average uncertainty of the RV data points is $\sim 0.00015~\mathrm{km~s}^{-1}$. 

We co-added the individual ESPRESSO spectra prior to carrying out the spectroscopic analysis presented in Sect.~\ref{stelpar}. To this aim, we Doppler-shifted the data to a common reference wavelength by cross-correlating the ESPRESSO spectra with the spectrum with the highest $S/N$. We finally performed a $S/N$-weighted co-addition of the Doppler-shifted spectra, while applying a sigma-clipping algorithm to remove possible cosmic-ray hits and outliers. The co-added spectrum has a $S/N$ of $\sim 900$ per pixel at 650\,nm.

\begin{table}
\centering
\caption{Log of ESPRESSO/VLT RV observations of HD\,22946.}
\label{espdatatable}
\begin{tabular}{ccc}
\hline
\hline
Time [$\mathrm{BJD_{TDB}}$]             & RV value [$\mathrm{km~s}^{-1}$]         & $\pm 1\sigma$ [$\mathrm{km~s}^{-1}$]\\
\hline
\hline
2458524.56069831                                & 16.85125                                      & 0.00011\\ 
2458525.55490396                                & 16.85217                                      & 0.00013\\
2458526.59541816                                & 16.85512                                      & 0.00011\\
2458527.63233315                                & 16.85284                                      & 0.00022\\
2458535.62345024                                & 16.84839                                      & 0.00036\\
2458540.53620531                                & 16.85020                                      & 0.00010\\
2458550.57174504                                & 16.85549                                      & 0.00020\\
2458552.56783808                                & 16.85330                                      & 0.00016\\
2458553.51738686                                & 16.86251                                      & 0.00011\\
2458556.50131285                                & 16.85536                                      & 0.00014\\
2458557.50492574                                & 16.85738                                      & 0.00009\\
2458557.56483059                                & 16.85716                                      & 0.00010\\
2458558.52709593                                & 16.85741                                      & 0.00010\\
2458559.54006749                                & 16.85690                                      & 0.00016\\
\hline
\hline
\end{tabular}
\end{table}

\section{Data analysis and first results}
\label{dataan}

\subsection{Stellar parameters}
\label{stelpar}

The spectroscopic stellar parameters (the effective temperature $T_{\mathrm{eff}}$, the surface gravity $\log g$, the microturbulent velocity $v_\mathrm{mic}$, and the metallicity [Fe/H]; see Table \ref{stelpartab}) were derived using the \texttt{ARES} and \texttt{MOOG} codes, following the same methodology as described in \citet{Sousa-21}, \citet{Sousa-14}, and \citet{Santos-13}. We used the latest version of the \texttt{ARES} code\footnote{The last version, \texttt{ARES} v2, can be downloaded at \url{https://github.com/sousasag/ARES}.} \citep{Sousa-07, Sousa-15} to measure the equivalent widths of iron lines on the combined ESPRESSO spectrum. We used a minimisation procedure to find ionisation and excitation equilibrium and converge to the best set of spectroscopic parameters. This procedure makes use of a grid of Kurucz model atmospheres \citep{Kurucz-93} and the radiative transfer code \texttt{MOOG} \citep{Sneden-73}.

To derive the radius of the host star HD\,22946, we used a Markov-Chain Monte Carlo (MCMC) modified infrared flux method. This enables us to calculate the bolometric flux using stellar atmospheric models defined by our spectral analysis to build spectral energy distributions (SEDs) that are compared with broadband fluxes and uncertainties from the most recent data releases for the following bandpasses: {\it Gaia} $G$, $G_\mathrm{BP}$, and $G_\mathrm{RP}$, 2MASS $J$, $H$, and $K$, and \textit{WISE} $W1$ and $W2$ \citep{Skrutskie2006,Wright2010,GaiaCollaboration2021}. From the bolometric flux, we then determine stellar effective temperature and angular diameter; this latter is converted to a radius using the offset-corrected \textit{Gaia} parallax \cite{Lindegren2021}. We used Bayesian modeling averaging of the \textsc{atlas} \citep{Kurucz1993,Castelli2003} and \textsc{phoenix} \citep{Allard2014} catalogues to produce a weighted averaged posterior distribution of the stellar radius in order to account for uncertainties in stellar atmospheric modelling. We find a value of $R_\mathrm{s}=1.117\pm0.009\, \mathrm{R}_\odot$, which is in $3\sigma$ agreement with the value of $1.157 \pm 0.025~\mathrm{R}_\odot$ presented by the discoverers.

We finally determined the stellar mass $M_\mathrm{s}$ and stellar age $t_\mathrm{s}$ using two different sets of stellar evolutionary models, namely PARSEC\footnote{\textsl{PA}dova and T\textsl{R}ieste \textsl{S}tellar \textsl{E}volutionary \textsl{C}ode: \url{http://stev.oapd.inaf.it/cgi-bin/cmd}} v1.2S \citep{marigo17} and CLES (Code Liègeois d'Évolution Stellaire), see \citet{scuflaire08}. More specifically, we employed the isochrone-placement algorithm developed by \citet{bonfanti15,bonfanti16} to interpolate the input parameters ($T_{\mathrm{eff}}$, [Fe/H], $R_\mathrm{s}$) within pre-computed grids of PARSEC v1.2S isochrones and tracks to derive a first pair of mass and age. A second pair of mass and age values, instead, was retrieved by inputting $T_{\mathrm{eff}}$, [Fe/H], and $R_\mathrm{s}$ directly in the CLES code, which generates the best-fit stellar evolutionary track following the Levenberg-Marquadt minimisation scheme, as described in \citet{salmon21}. After carefully checking the mutual consistency of the two respective pairs of outcomes through the $\chi^2$-based methodology presented in \citet{bonfanti21}, we finally merged (i.e. summed) the two $M_\mathrm{s}$ and $t_\mathrm{s}$ results and obtained $M_\mathrm{s}=1.098 \pm 0.040~\mathrm{M}_\odot$ and $t_\mathrm{s}=2.5 \pm 1.0$ Gyr. The mass parameter value of the host star agrees within the uncertainty with the  value provided in the discovery paper, which is $1.104 \pm 0.012~\mathrm{M}_\odot$. However, the planet host seems to be younger than previously presented by C22. The discoverers obtained a value of $5.0 \pm 1.0$ Gyr. More parameter values, including from this work, are compared with the discovery-paper parameter values in Table \ref{stelpartab}.   

\begin{table}
\centering
\caption{Fundamental parameters of the exoplanet host HD\,22946.}
\label{stelpartab}
\begin{tabular}{lll}
\hline
\hline
Parameter [unit]                                & Value                                                 & Source\\
\hline
\hline
Name                                                    & HD\,22946                                      & --\\
TOI ID                                                  & 411                                                    & G2021\\
TIC ID                                          & 100990000                                      & S2018\\
\textit{Gaia} DR3 ID                            & 4848767461548943104            & G2022\\
RA (J2016) [deg]                                & $54.819528$                                    & G2022\\
Dec (J2016) [deg]                               & $-42.76304$                                    & G2022\\
$T$ (TESS) [mag]                                & $7.757 \pm 0.006$                     & S2018\\
$G$ (\textit{Gaia}) [mag]                       & $8.13 \pm 0.69$                               & G2022\\
$J$ [mag]                                               & $7.250 \pm 0.027$                         & C2003\\
$H$ [mag]                                       & $7.040 \pm 0.044$                     & C2003\\
$K$ [mag]                                       & $6.981 \pm 0.029$                     & C2003\\
$T_\mathrm{eff}$ [K]                            & $6040 \pm 48$                                  & C2022\\
$T_\mathrm{eff}$ [K]                            & $6169 \pm 64$                                  & This work\\
$R_\mathrm{s}$ [$R_\odot$]              & $1.157 \pm 0.025$                     & C2022\\
$R_\mathrm{s}$ [$R_\odot$]              & $1.117 \pm 0.009$                     & This work\\
$M_\mathrm{s}$ [$M_\odot$]              & $1.104 \pm 0.012$                                     & C2022\\
$M_\mathrm{s}$ [$M_\odot$]              & $1.098^{+0.040}_{-0.039}$             & This work\\
$\log g$ [cgs]                                  & $4.26 \pm 0.15$                               & C2022\\
$\log g$ [cgs]                                  & $4.47 \pm 0.11$                               & This work\\
$[\mathrm{Fe/H}]$ [dex]                 & $-0.14 \pm 0.07$                              & C2022\\
$[\mathrm{Fe/H}]$ [dex]                 & $-0.08 \pm 0.04$                              & This work\\
$t_\mathrm{s}$ [Gyr]                            & $5.0 \pm 1.0$                          & C2022\\
$t_\mathrm{s}$ [Gyr]                            & $2.5 \pm 1.0$                          & This work\\
$v_\mathrm{mic}$ [km~s$^{-1}$]  & $1.25 \pm 0.03$                               & This work\\ 
\hline
\hline
\end{tabular}
\tablefoot{Abbreviations refer to the following sources: G2021 = \citet{Guerrero1}, S2018 = \citet{Stassun1}, G2022 = \citet{GaiaCollaboration2022}, C2003 = \citet{Cutri1}, C2022 = \citet{Cacci2}.}
\end{table}

\subsection{Period aliases of HD\,22946d from the TESS data}
\label{dataanper}

In order to determine each possible period alias and to schedule CHEOPS observations of planet d, we first performed a period analysis of the available TESS data. For this purpose, we used the \texttt{MonoTools} package\footnote{See \url{https://github.com/hposborn/MonoTools}.} \citep{Osborn1}, which is able to model transit light curves in case of multiple transits, duotransits, and monotransits, as well as multiple systems with combinations of such candidates, with both radial velocities and transit photometry. The package calculates a marginalised probability distribution across all allowed aliases for a given transit model by combining priors for each alias. The probabilities are estimated based on two major assumptions, namely that short-period orbits are highly favoured over long-period ones due to a combination of geometric probability and window function, and that planets in multi-planet systems have low eccentricities \citep{Kipping1, Kipping2, Eylen1}. More details about this software can be found in \citet{Osborn1}.

The TESS data described in Sect. \ref{tessdata} were used during the fitting procedure using \texttt{MonoTools}. In the case of planet b, we set as input parameters the reference mid-transit time of $T_\mathrm{c} = 2~458~385.7318$ $\mathrm{BJD}_\mathrm{TDB}$, the orbital period of $P_\mathrm{orb} = 4.040330 \pm 0.000010$ d, the transit duration (transit width) of $W = 3.4$ hr, and the transit depth of $D = 134$ ppm. In the case of planet c, the inputs were $T_\mathrm{c} = 2~458~386.1878$ $\mathrm{BJD}_\mathrm{TDB}$, $P_\mathrm{orb} = 9.573117 \pm 0.000020$ d, $W = 3.8$ hr, and $D = 389$ ppm. For planet d, we set as input parameters the two mid-transit times detected by TESS, namely $t_\mathrm{tr,1} = 2~458~425.1657$ $\mathrm{BJD}_\mathrm{TDB}$ and $t_\mathrm{tr,2} = 2~459~136.5357$ $\mathrm{BJD}_\mathrm{TDB}$, the transit duration of $W = 6.5$ hr and the transit depth of $D = 478$ ppm. These parameters were calculated from the TESS data alone.     

The orbital period aliases of planet d with a probability of $p > 1\%$ are listed in Table \ref{tessperaliases}. The software \texttt{MonoTools} forecasted that a transit of planet d with the orbital period alias number 2 would take place on 25 October 2021, with a mid-transit time of $T_\mathrm{c} = 2~459~513.1441$ $\mathrm{BJD}_\mathrm{TDB}$. This forecasted event was observed during the third CHEOPS  visit (see Table \ref{cheopsobslog}), but the expected transit of planet d did not happen; only the transit of planet b was recorded that time. After this observation, we were able to exclude the period alias of $P = 41.8454$ d from the list of possible aliases. The next forecast predicted a transit of planet d on 28 October 2021, with a mid-transit time of $T_\mathrm{c} = 2~459~515.9338$ $\mathrm{BJD}_\mathrm{TDB}$, which means that, in this case, the alias number 4 (see Table \ref{tessperaliases}) was preferred as its true orbital period. This forecasted event was observed with CHEOPS during its fourth visit. This time, the transit of planet d was successfully detected together with a transit of planet c, confirming that the period alias of $P = 47.4248$ d is the true orbital period of planet d. This result also confirms that the second transit-like feature of planet d, observed by TESS in sector number 30, was a real transit event and not an instrumental artifact as considered by C22. Alternatively, the dip observed at $2~459~136.5357~\mathrm{BJD}_\mathrm{TDB}$ was a mixture of instrumental effects and the transit of planet d. With this gathered knowledge about the true orbital period of planet d, we were able to combine CHEOPS and TESS photometric observations and RV measurements in order to improve the orbital and planetary parameters of the HD\,22946 system, which were previously obtained only from the TESS and RV data by the discoverers.                       

 \begin{table}
 \centering
 \caption{Orbital period aliases of the planet HD\,22946d.}
 \label{tessperaliases}
 \begin{tabular}{ccc}
 \hline
 \hline
 Alias          & Period alias ($P$)            & Probability ($p$)\\
 No.            & [d]                                   & [\%]\\   
 \hline
 \hline
 1              & 39.5206                       & $17.420$\\
 2              & 41.8454                       & $20.078$\\
 3              & 44.4607                       & $20.341$\\
 4              & 47.4248                       & $18.113$\\ 
 5              & 50.8122                       & $13.445$\\
 6              & 54.7209                       & $7.061$\\
 7              & 59.2809                       & $2.756$\\
 8              & 64.6701                       & $\sim 1.0$\\
 \hline
 \hline
 \end{tabular}
 \tablefoot{Only the period aliases with a probability of $p > 1\%$ are listed here, as calculated by the \texttt{MonoTools} package from TESS data alone, i.e. before CHEOPS observations.}
 \end{table}

\begin{table*}
\footnotesize
\centering
\caption{Best-fitting and derived system and planetary parameters of the HD\,22946 planetary system.}
\label{finalparams}
\begin{tabular}{lllll}
\hline
\hline
Parameter [unit]                                                                        & Description                                                             & HD\,22946b                                                              & HD\,22946c                                                                              & HD\,22946d\\   
\hline
\hline
$T_\mathrm{c}$ [$\mathrm{BJD}_\mathrm{TDB}$]                    & reference mid-transit time                                        & $2~458~385.7321^{+0.0022}_{-0.0031}$                 & $2~459~161.60861^{+0.00069}_{-0.00072}$                        & $2~459~136.53720^{+0.00087}_{-0.00083}$\\
$P_\mathrm{orb}$ [d]                                                             & orbital period                                                                & $4.040295^{+0.000015}_{-0.000014}$              & $9.573083 \pm 0.000014$                                                         & $47.42489^{+0.00010}_{-0.00011}$\\
$b$                                                                                              & impact parameter                                                      & $0.21^{+0.11}_{-0.13}$                                          & $0.504^{+0.024}_{-0.026}$                                                         & $0.456^{+0.026}_{-0.028}$\\
$a/R_\mathrm{s}$                                                                         & scaled semi-major axis                                                & $11.03 \pm 0.12$                                                        & $19.61^{+0.22}_{-0.23}$                                                         & $57.00 \pm 0.66$\\
$a$ [au]                                                                                        & semi-major axis                                                 & $0.05727^{+0.00085}_{-0.00082}$                         & $0.1017^{+0.0015}_{-0.0014}$                                                   & $0.2958^{+0.0044}_{-0.0042}$\\
$R_\mathrm{p}/R_\mathrm{s}$                                                      & planet-to-star radius ratio                                   & $0.01119^{+0.00031}_{-0.00032}$                         & $0.01912^{+0.00026}_{-0.00027}$                                                & $0.02141^{+0.00046}_{-0.00045}$\\
$t_\mathrm{dur}$ [d]                                                                    & transit duration                                                        & $0.1281^{+0.0026}_{-0.0037}$                            & $0.1535^{+0.0015}_{-0.0014}$                                                 & $0.2701^{+0.0019}_{-0.0020}$\\ 
$R_\mathrm{p}$ [$\mathrm{R}_\oplus$]                                    & planet radius                                                           & $1.362 \pm 0.040$                                               & $2.328^{+0.038}_{-0.039}$                                                         & $2.607 \pm 0.060$\\
$M_\mathrm{p}$ [$\mathrm{M}_\oplus$]                                    & planet mass$^{\star}$                                                   & $13.71$                                                                         & $9.72$                                                                                          & $26.57$\\ 
$M_\mathrm{p,est}$ [$\mathrm{M}_\oplus$]                                & estimated planet mass$^{\clubsuit}$                     & $2.42 \pm 0.12$                                                 & $6.04 \pm 0.17$                                                                        & $7.32 \pm 0.28$\\
$M_\mathrm{p,est}$ [$\mathrm{M}_\oplus$]                                & estimated planet mass$^{\heartsuit}$                    & $2.61 \pm 0.27^{\dagger}$                                         & $6.61 \pm 0.17^{\ddagger}$                                                     & $7.90 \pm 0.28^{\ddagger}$\\   
$\rho_\mathrm{p}$ [$\mathrm{g~cm}^{-3}$]                                & planet density$^{\star}$                                                & $18.96$                                                                         & $3.15$                                                                                  & $10.80$\\
$K$ [$\mathrm{m~s}^{-1}$]                                                       & RV semi-amplitude$^{\star}$                                     & $5.05$                                                                         & $2.70$                                                                                  & $4.31$\\
$I_\mathrm{p}$ [$\mathrm{W~m}^{-2}$]                                    & insolation flux                                                         & $673~884^{+32~444}_{-31~606}$                           & $213~337^{+10~270}_{-10~006}$                                         & $25~261^{+1216}_{-1184}$\\ 
$T_\mathrm{surf}$ [K]                                                           & surface temperature$^{\diamond}$                                & $1241 \pm 14$                                                     & $931 \pm 11$                                                                           & $546 \pm 6$\\
TSM                                                                                             & transmission spectroscopy metric$^{\sharp}$     & $43 \pm 4$                                                             & $63 \pm 2$                                                                             & $43 \pm 2$\\    
\hline
\hline
\end{tabular}
\tablefoot{Based on the joint fit of the TESS and CHEOPS photometric data and RV observations. Table shows the best-fitting value of the given parameter and its $\pm 1\sigma$ uncertainty. The fitted values correspond to quantile 0.50 (median) and the uncertainties to quantils $\pm 0.341$ in the parameter distributions obtained from the samples. $^{\star}$Only the $3\sigma$ upper limits are listed here due to the low number of the RV observations. $^{\diamond}$Assuming the albedo value of 0.2. $^{\clubsuit}$Based on the relations presented by \citet{Chen1}, assuming that the planet radii are from the interval of $1.23 < R_\mathrm{p} < 14.26~\mathrm{R}_\oplus$. $^{\heartsuit}$Based on the relations presented by \citet{Otegi1}. $^{\dagger}$Assuming that $\rho_\mathrm{p} > 3.3~\mathrm{g~cm}^{-3}$. $^{\ddagger}$Assuming that $\rho_\mathrm{p} < 3.3~\mathrm{g~cm}^{-3}$. $^{\sharp}$Based on the criteria set by \citet{Kempton1}, using the scale factor of 1.26, the stellar radius of $R_\mathrm{s} = 1.117 \pm 0.009~\mathrm{R_\odot}$ and $J = 7.250 \pm 0.027$ mag (see Table \ref{stelpartab}).}
\end{table*}

\subsection{CHEOPS, TESS, and RV combined model}
\label{dataancheopstess}

In order to produce accurate planetary parameters for all three planets, we built a combined model using all available data, that is, TESS photometry (described in Sect. \ref{tessdata}), CHEOPS photometry (described in Sect. \ref{cheopsdata}), and ESPRESSO RVs (described in Sect. \ref{espdata}). The combined model was built using the \texttt{PyMC3} package\footnote{See \url{https://pypi.org/project/pymc3/}.} \citep{Salvatier1}, which performs Hamiltonian Monte Carlo (HMC) sampling, with Keplerian orbits modeled with \texttt{exoplanet} package\footnote{See \url{https://pypi.org/project/exoplanet/}.} \citep{Foreman1}. We used Gaussian processes (GPs) to model the stellar variability present in the TESS light curve, opting for a simple harmonic oscillator (SHO) kernel implemented in the \texttt{celerite} package \citep{Foreman2} and a quality factor $Q=1/\sqrt{2,}$ as is common for quasi-periodic stellar variability. In order to speed up sampling, we binned the TESS data to 30 minute bins far from transits, keeping 2 minute data near transit. As we have reasonable prior knowledge from theoretical analyses for the expected stellar limb-darkening (LD) parameters for HD\,22946, we used these as priors in the analysis. We used the quadratic LD law and interpolated tables of coefficients calculated for the TESS \citep{Claret1} and CHEOPS \citep{Claret2} passbands using the derived stellar parameters of $T_\mathrm{eff} = 6169$ K and $\log g = 4.47$ (cgs). In order to guard against systematic errors, we inflated the $\sigma$ for each parameter prior to 0.1.

Even though the \texttt{PIPE} light curves for HD\,22946 have fewer systematic features than the DRP light curves, they can still include flux variations due to the influence of various external factors. Therefore, we can improve the light curve by decorrelating the flux data against metadata generated for the instrument and target. To decipher which decorrelation vectors provide improvement, we ran an initial \texttt{PyMC3} model for each CHEOPS visit using all available ancillary data -- $\sin$ and $\cos$ of rollangle, background flux, $x$ and $y$ centroid positions, onboard temperature and time (which also fits short-timescale stellar variability). These parameters are normalised to have $\mu=0.0$ and $\sigma=1.0$, and decorrelation parameters are given normal priors with $\mu=0.0$ and $\sigma$ set by the root-mean-square (RMS) noise for each CHEOPS visit. For each visit model, we also included parameters for any planetary transits present in order to ensure the transits would not bias the model. After HMC sampling, we assessed each decorrelation parameter using the average and standard deviations, keeping only those parameters with a Bayes Factor of BF > 1. Despite this detrending, shorter-timescale variation can also be present as a function of roll angle ($\varphi$). Pure detrending against $\sin$ and $\cos$ of roll angle removes the largest amplitude systematic trends at low frequencies. These are those closest in timescale to the transit feature, and so a simpler detrending technique for such timescales guards against over-fitting of the transit. However, the CHEOPS light curve typically also contains systematic noise correlated with roll angle that is at a lower amplitude and higher frequency. This is not therefore adequately removed by simple $\sin$ and $\cos$ decorrelation. It is this noise that a more flexible GP is better able to model. We therefore also included a GP to model the variation of flux with roll-angle effects. To do this, we first found any potential large jumps in $\varphi$ and made sure the time series was continuous between these jumps (i.e. by moving the zero point and `wrapping around'). We then transformed the input data such that it is continuous in $x$ ---by sorting by $\varphi$ rather than time. Once again, we used a SHO kernel from \texttt{celerite} with quality factor $Q$ set at $1/\sqrt{2}$. As we expected the morphology of the variations to be preserved for all CHEOPS visits, we used a single shared kernel. We found that the linear decorrelation is the most important, decreasing the $\log$ likelihood by a factor of 1400, but the GP is responsible for a reduction of a further 450, which means that use of a GP to model roll-angle flux behaviour is well justified.

As multi-planet systems typically have low eccentricities $e$ \citep{van2019orbital}, and we lack the high number of RVs capable of resolving any differences in $e$, we chose to fit only circular orbits. In order to guard against unphysical negative values, we used broad log-normal priors for the key transit and RV amplitude parameters, that is, for $R_\mathrm{p}/R_\mathrm{s}$ (planet-to-star radius ratio) and $K$ (RV semi-amplitude). The quantities derived in Sect. \ref{stelpar} are used as priors on the stellar parameters in the model. For all datasets ---CHEOPS, TESS, and ESPRESSO---, we included a jitter term using a wide log-normal prior. We then sampled the combined model using the \texttt{PyMC\_ext 'sample'} function, which is specifically written for astrophysical applications, and allows us to group independent dataset parameters (e.g. the CHEOPS visit-specific decorrelation parameters) together, thereby speeding up sampling greatly. We used ten chains, tuning each for 1300 steps before sampling for a further 1800, resulting in 18~000 unique samples. The sample have effective sample sizes in the thousands, and the gelmin-rubin statistics are below 1.01 for all parameters, suggesting they are sufficiently uncorrelated and unbiased. The full list of fitted GP hyperparameters and detrending parameters with the corresponding best-fitting values can be found in Appendix \ref{hyppar}. The best-fitting and derived parameters of the system are described and discussed in Sect. \ref{res}.

\subsection{Search for transit-timing variations} 
\label{ttvsec}

\begin{figure}
\includegraphics[width=\columnwidth]{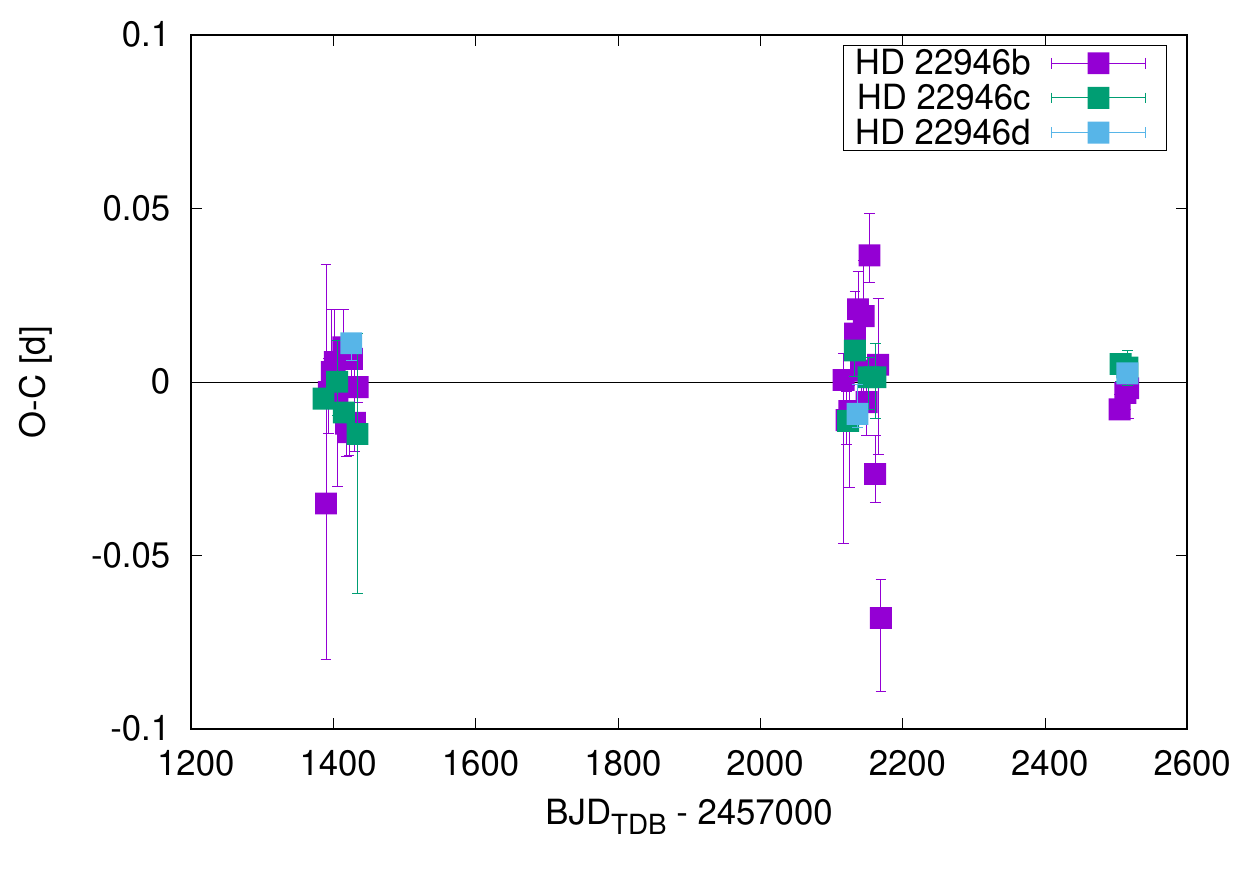}
\caption{Observed-minus-calculated (O-C) diagram of mid-transit times of the planets HD\,22946b, HD\,22946c, and HD\,22946d obtained using the \texttt{Allesfitter} package. The O-C values are consistent with a linear ephemeris, which means no significant TTVs are present in the system.}
\label{ocgraph} 
\end{figure}

In order to look for potential TTVs, we also ran a combined model using unconstrained timing for each planetary transit thanks to the \texttt{TTVorbit} function of \texttt{exoplanet}, and an independent analysis using the \texttt{Allesfitter} software\footnote{See \url{https://www.allesfitter.com/home}.} \citep{allesfitter-code, allesfitter-paper}, applying a nested sampling fit. Although C22 already performed such an analysis and found no obvious sign of TTVs in the system, we repeated this procedure, but in this case using the CHEOPS data as well. This means mainly that we included three transits of planet d in the analysis and used a longer time baseline. We used the same dataset as in Sect. \ref{dataancheopstess}, which was co-fitted with a GP using the \texttt{celerite SHO} kernel in both cases. All planetary and system parameters were fixed as derived previously, only the GP hyperparameters, the detrending parameters, and the observed-minus-calculated (O-C) parameters for individual mid-transit times were fitted. Both solutions are consistent with a linear ephemeris, which means we did not find any indication of a quadratic trend in the data, in agreement with the conclusion made by the discoverers. As an illustration, the obtained O-C diagram of the mid-transit times for planets b, c, and d from the \texttt{Allesfitter} package is depicted in Fig. \ref{ocgraph}. We can see that the O-C values are scattered around O-C = 0.0 d, which means that no significant TTVs are present in the system.    

\section{Final results and discussion}
\label{res}

\begin{figure*}
\centering
\centerline{
\includegraphics[width=\columnwidth]{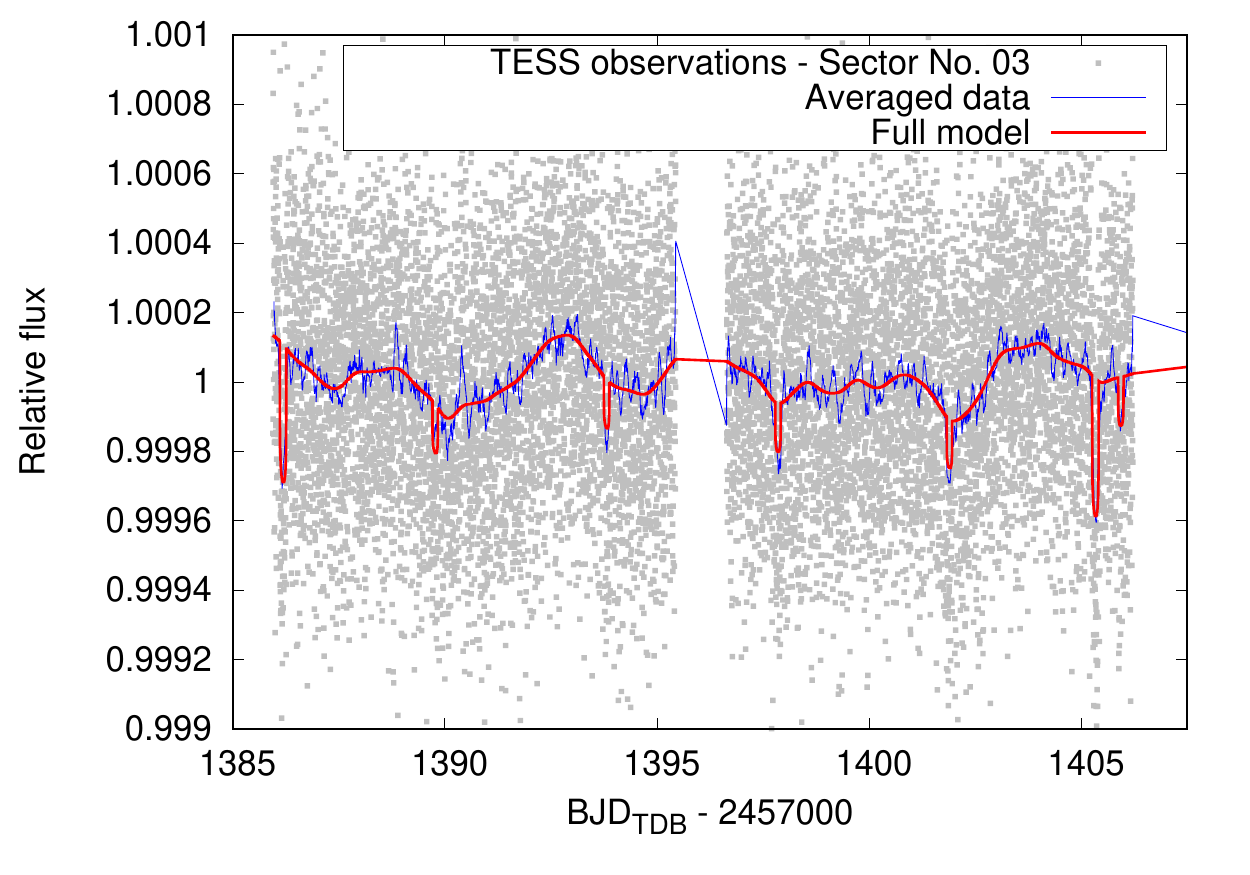}
\includegraphics[width=\columnwidth]{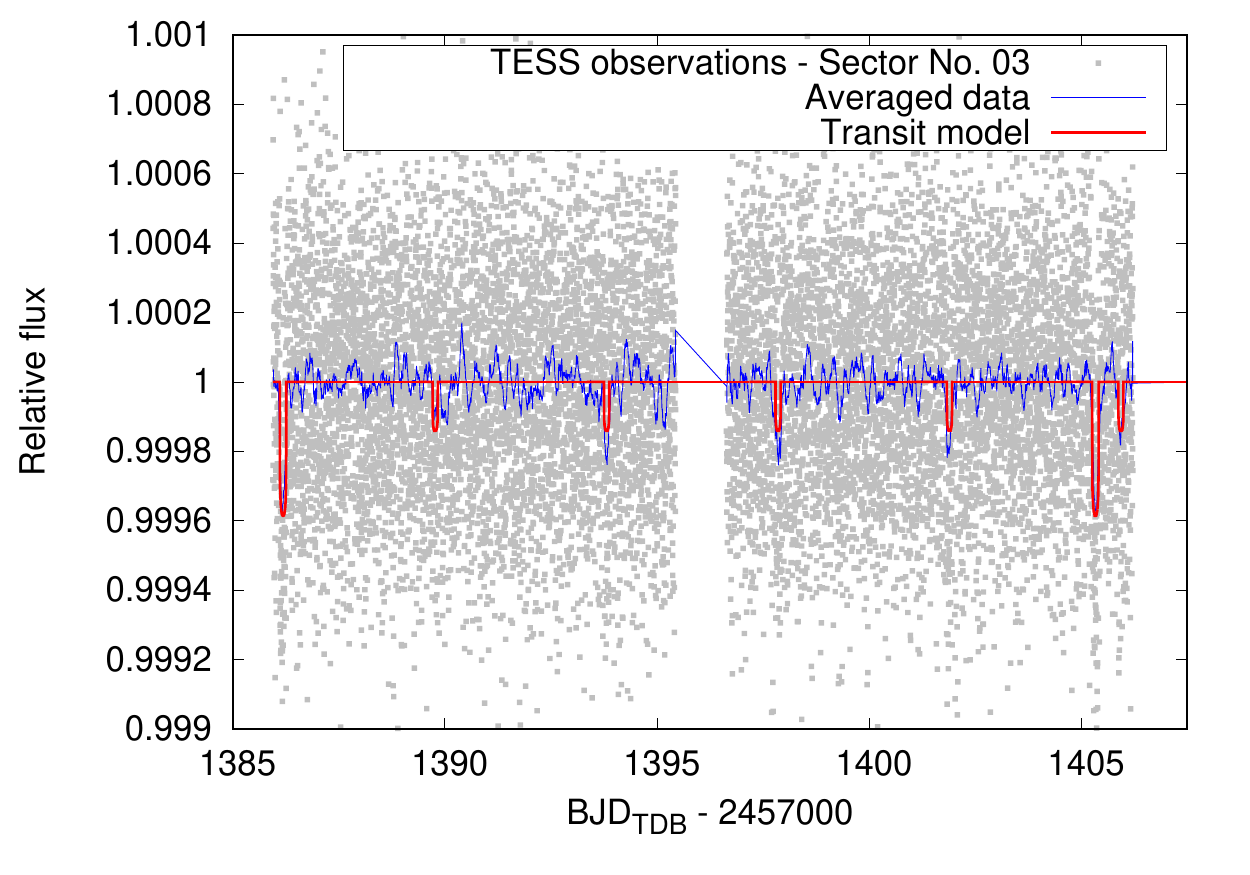}}
\centerline{
\includegraphics[width=\columnwidth]{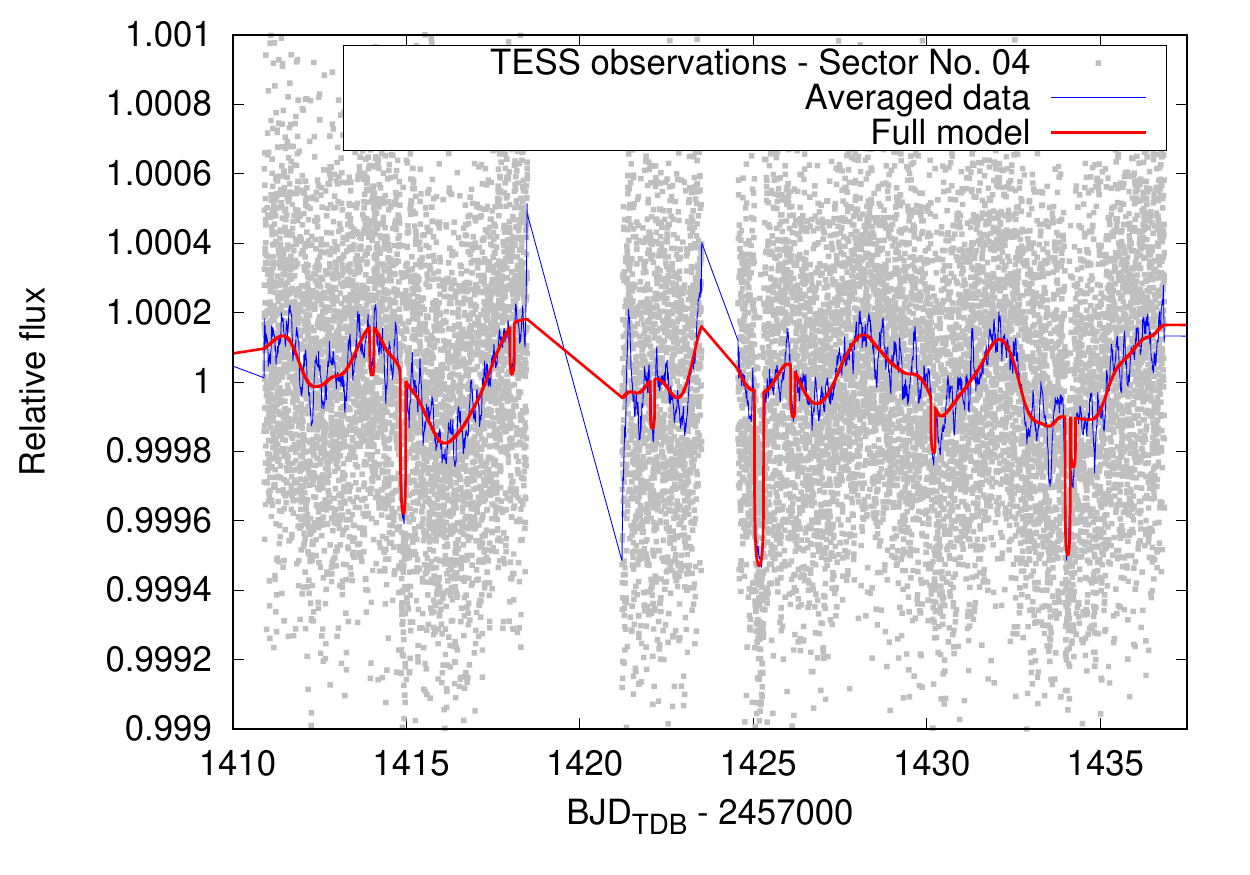}
\includegraphics[width=\columnwidth]{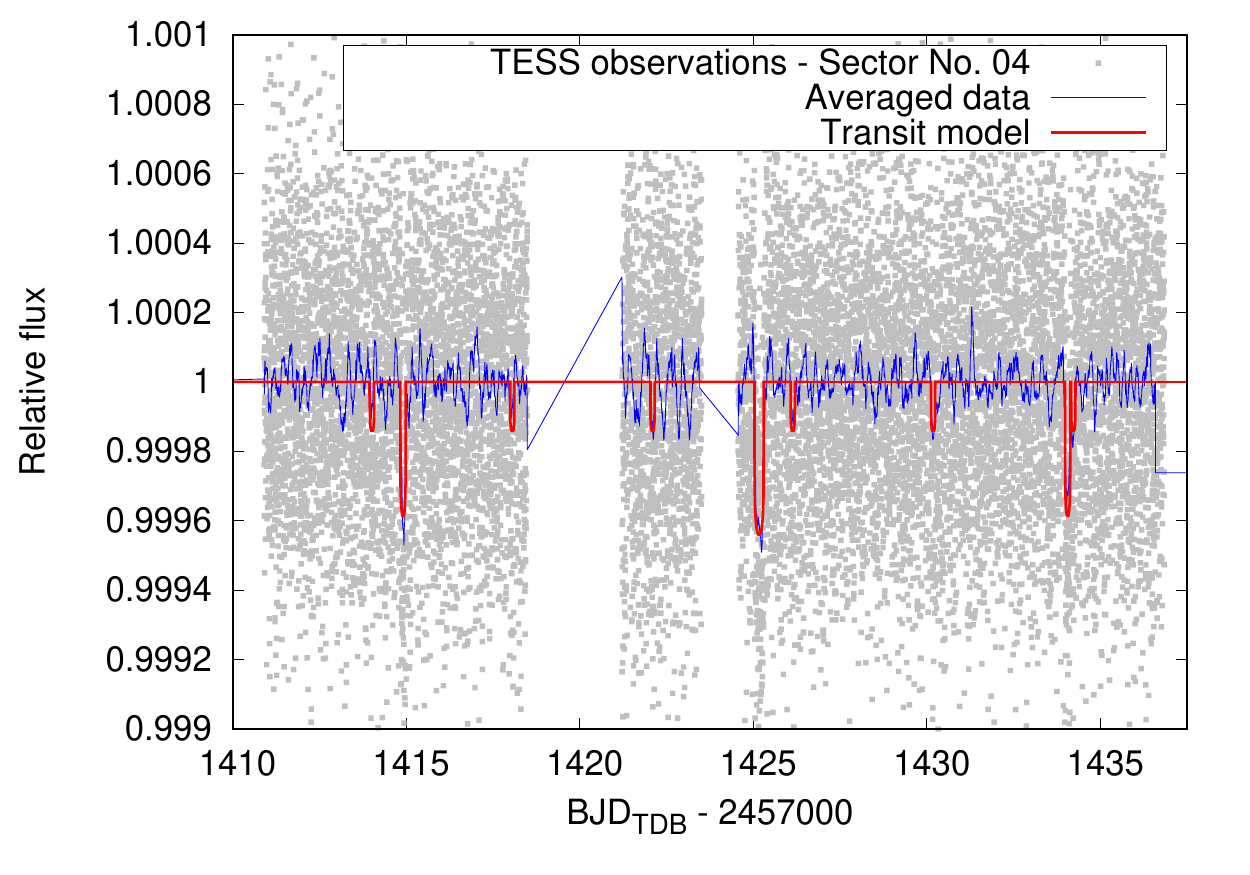}}
\centerline{
\includegraphics[width=\columnwidth]{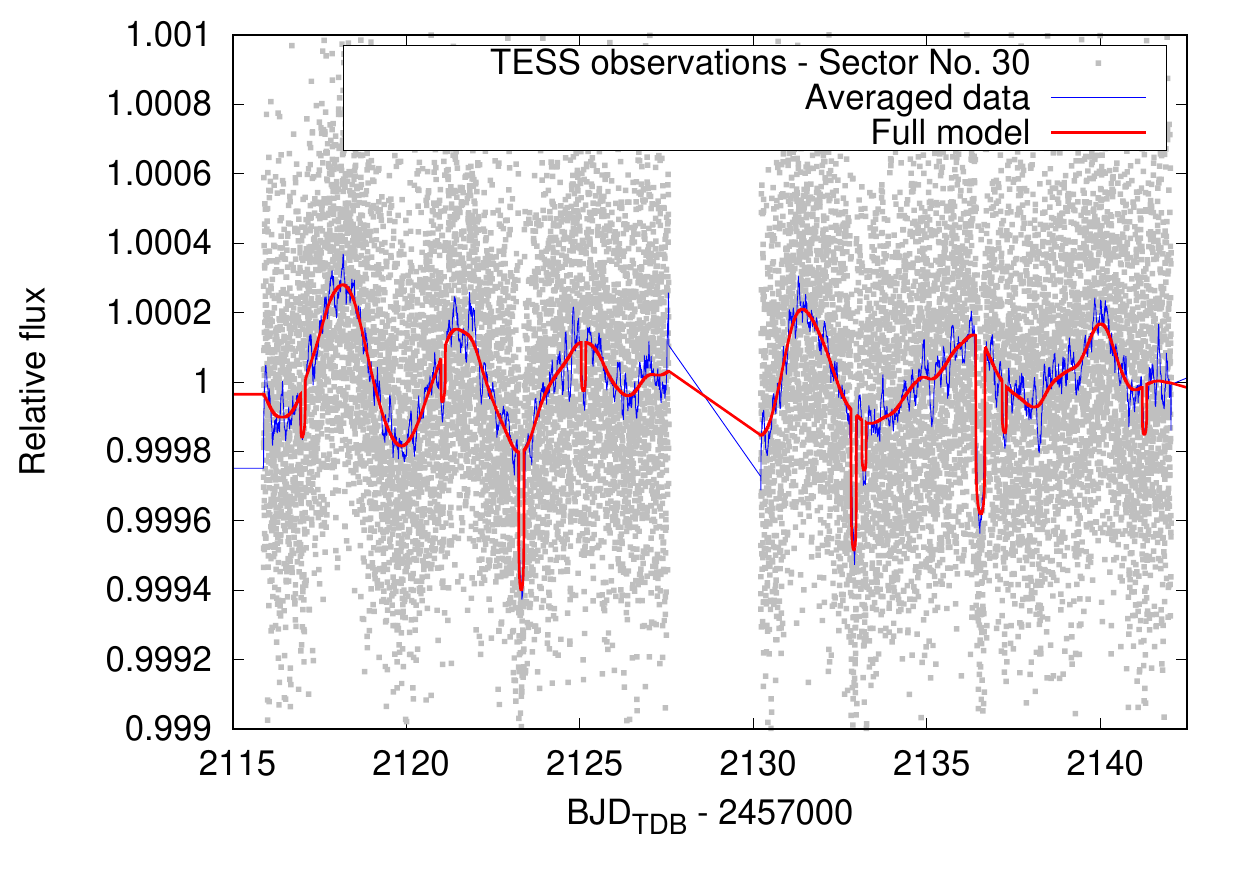}
\includegraphics[width=\columnwidth]{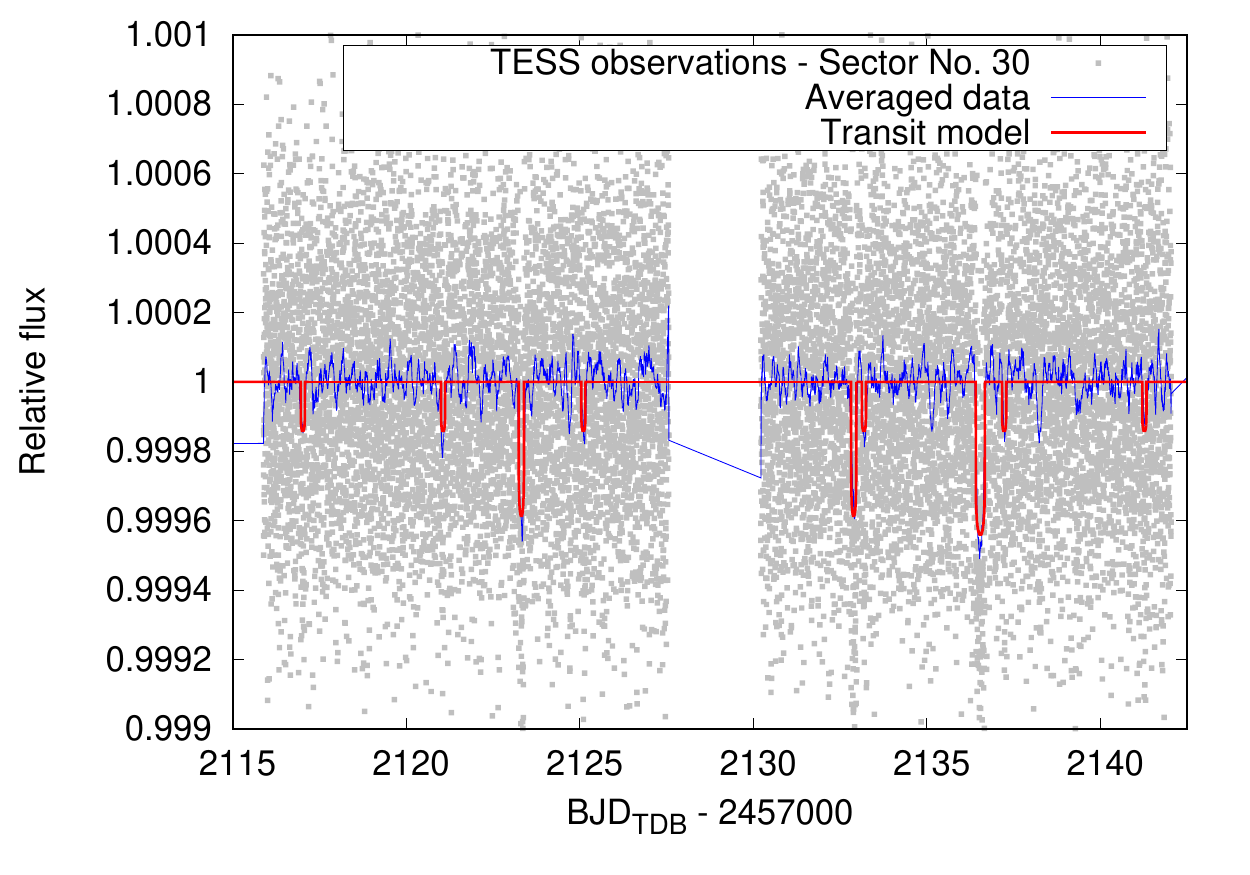}}
\caption{TESS observations of the transiting planets HD\,22946b, HD\,22946c, and HD\,22946d from sector numbers 3, 4, and 30 overplotted with the best-fitting model. This model was derived based on the entire CHEOPS and TESS photometric dataset and the RV observations from ESPRESSO via joint analysis of the data. The left-hand panels show the non-detrended data overplotted with the full model, while the right-hand panels show the detrended data overplotted with the transit model. We averaged the TESS data for better visualisation of the transit events using a running average with steps and width of 0.009 and 0.09 d, respectively. We note that an interruption in communications between the instrument and spacecraft occurred  at $2~458~418.54~\mathrm{BJD}_\mathrm{TDB}$, resulting in an instrument turn-off until $2~458~421.21~\mathrm{BJD}_\mathrm{TDB}$. No data were collected during this period.}
\label{tessdataplot1} 
\end{figure*}

\begin{figure*}
\centering
\centerline{
\includegraphics[width=\columnwidth]{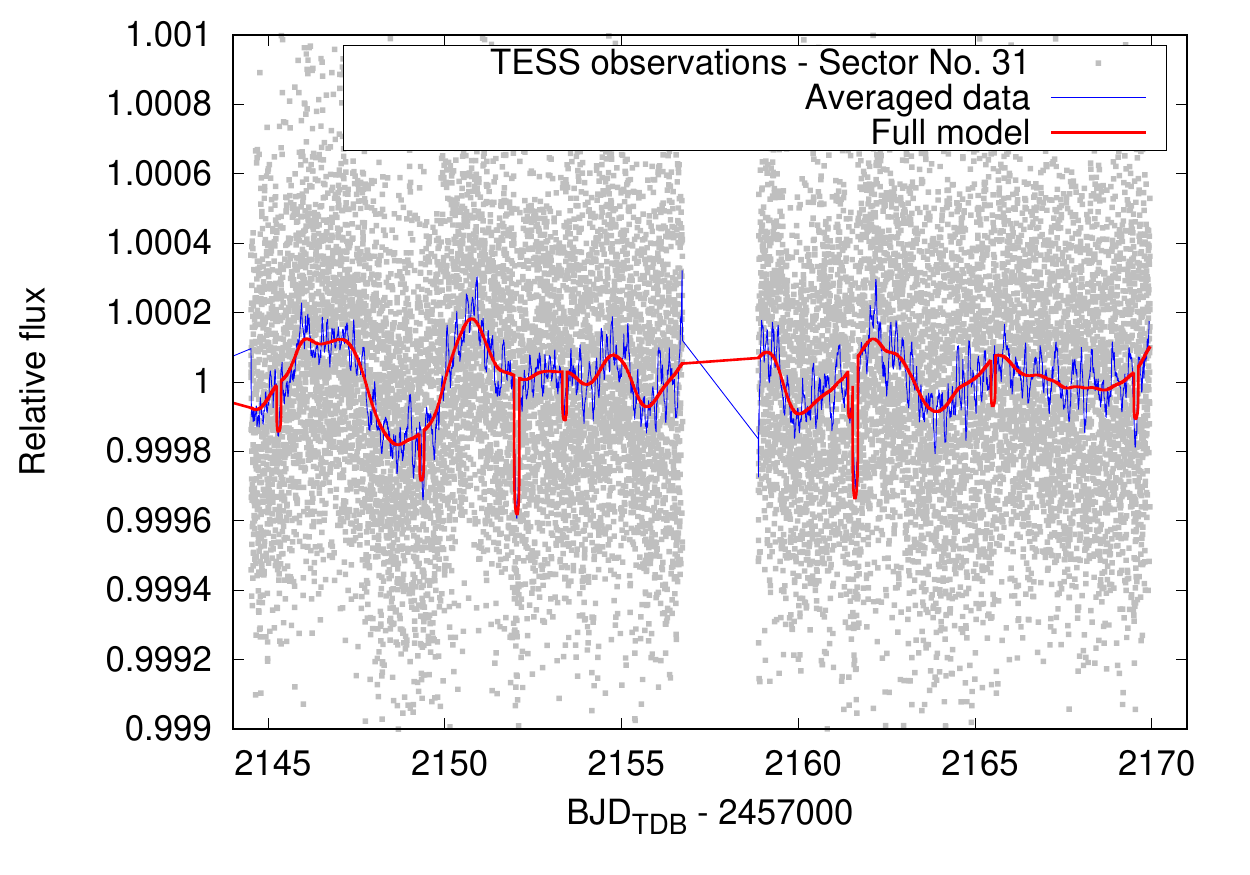}
\includegraphics[width=\columnwidth]{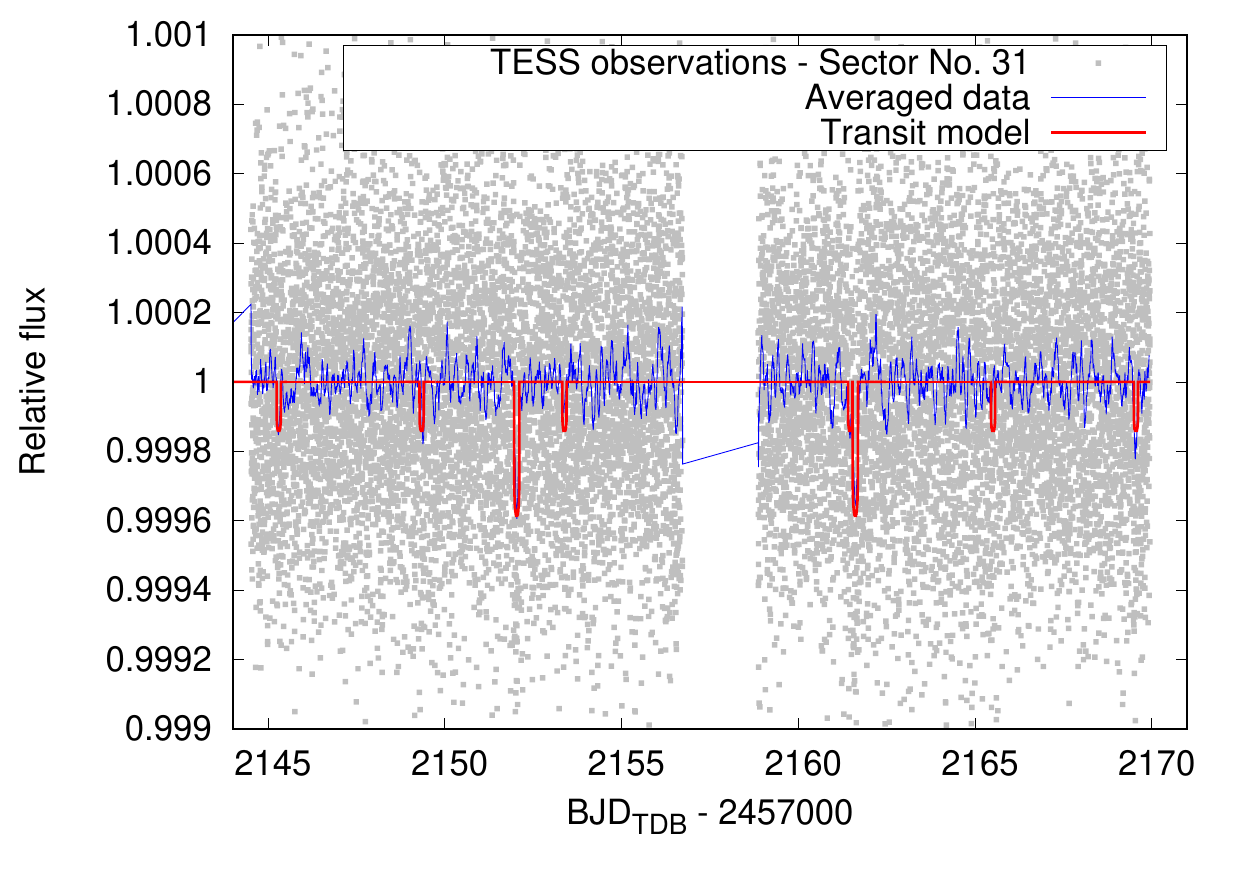}}
\caption{As in Fig. \ref{tessdataplot1}, but for the TESS sector number 31.}
\label{tessdataplot2} 
\end{figure*}

\begin{figure*}
\centering
\centerline{
\includegraphics[width=\columnwidth]{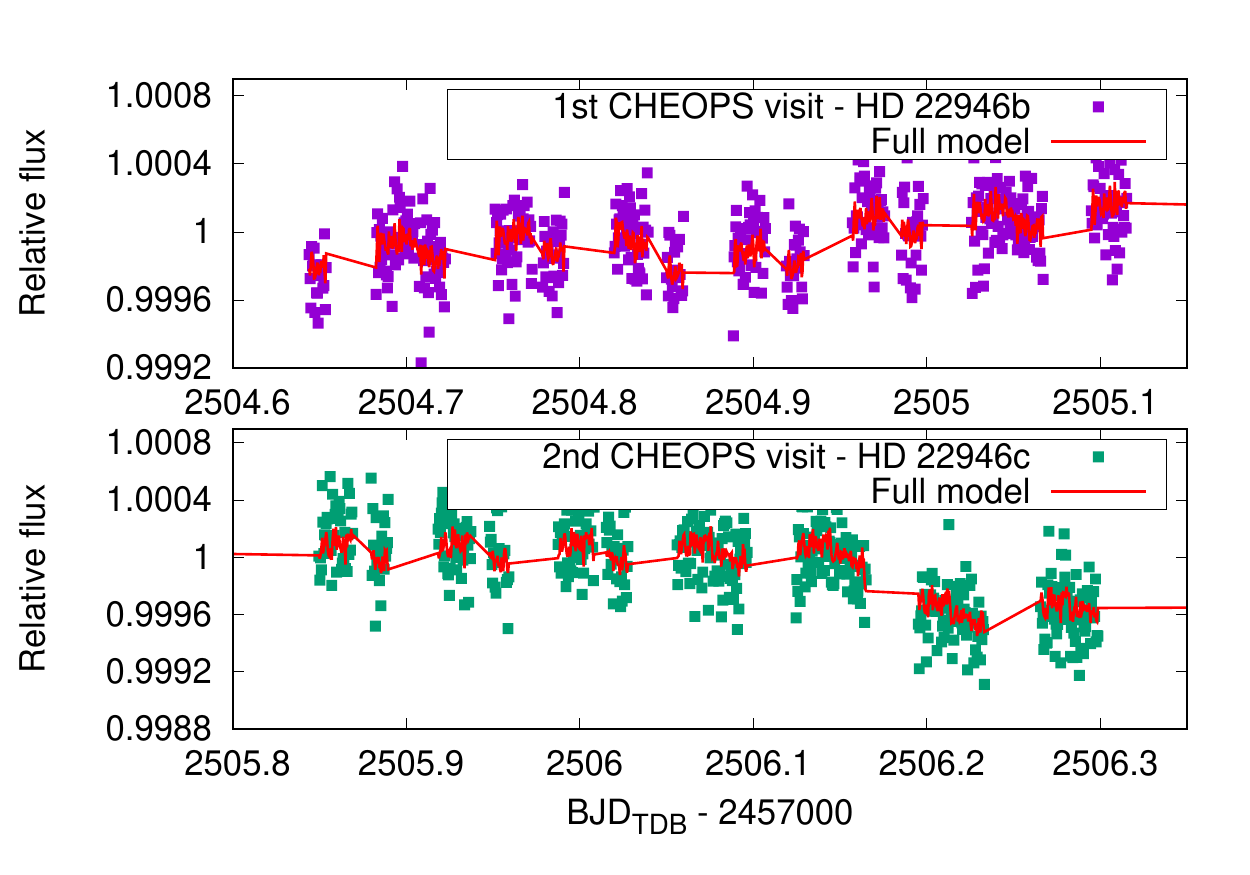}
\includegraphics[width=\columnwidth]{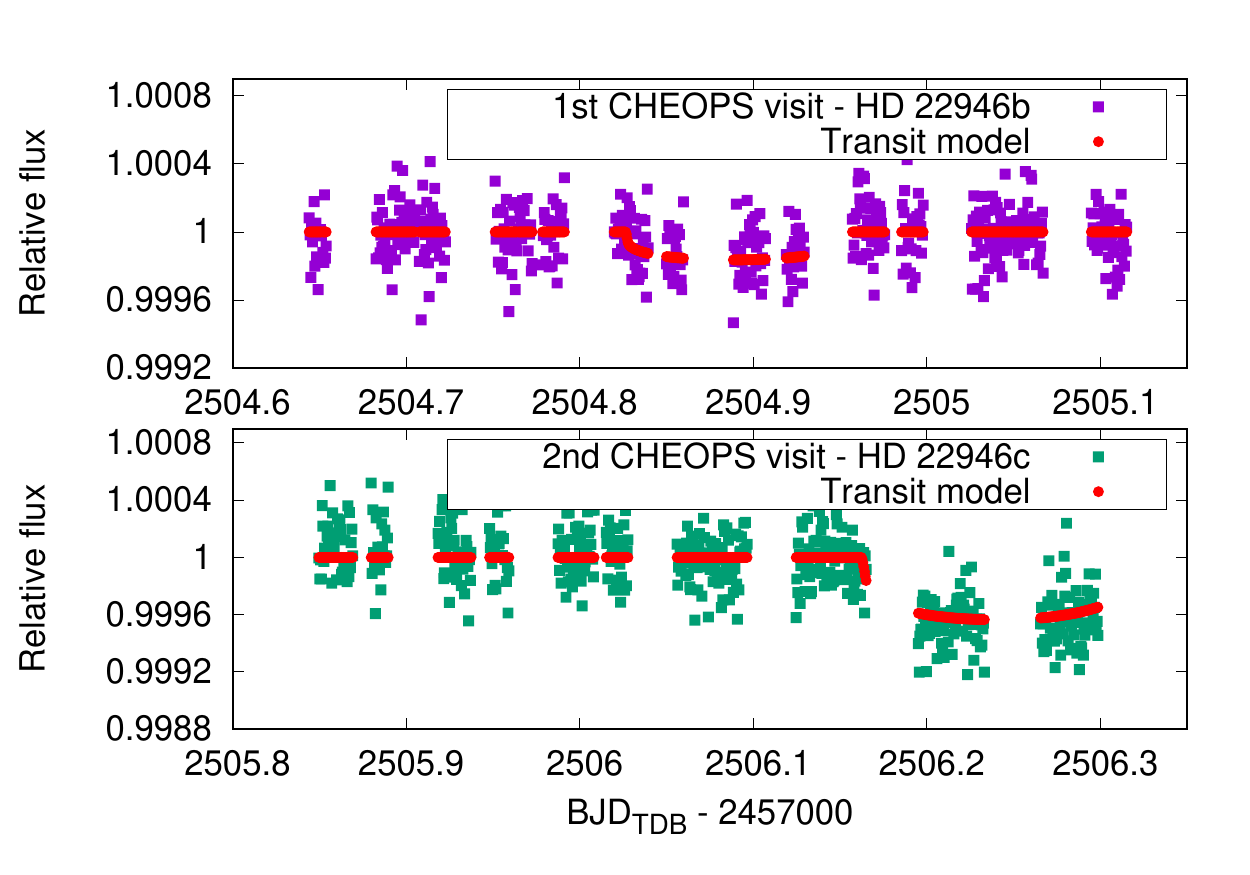}}
\centerline{
\includegraphics[width=\columnwidth]{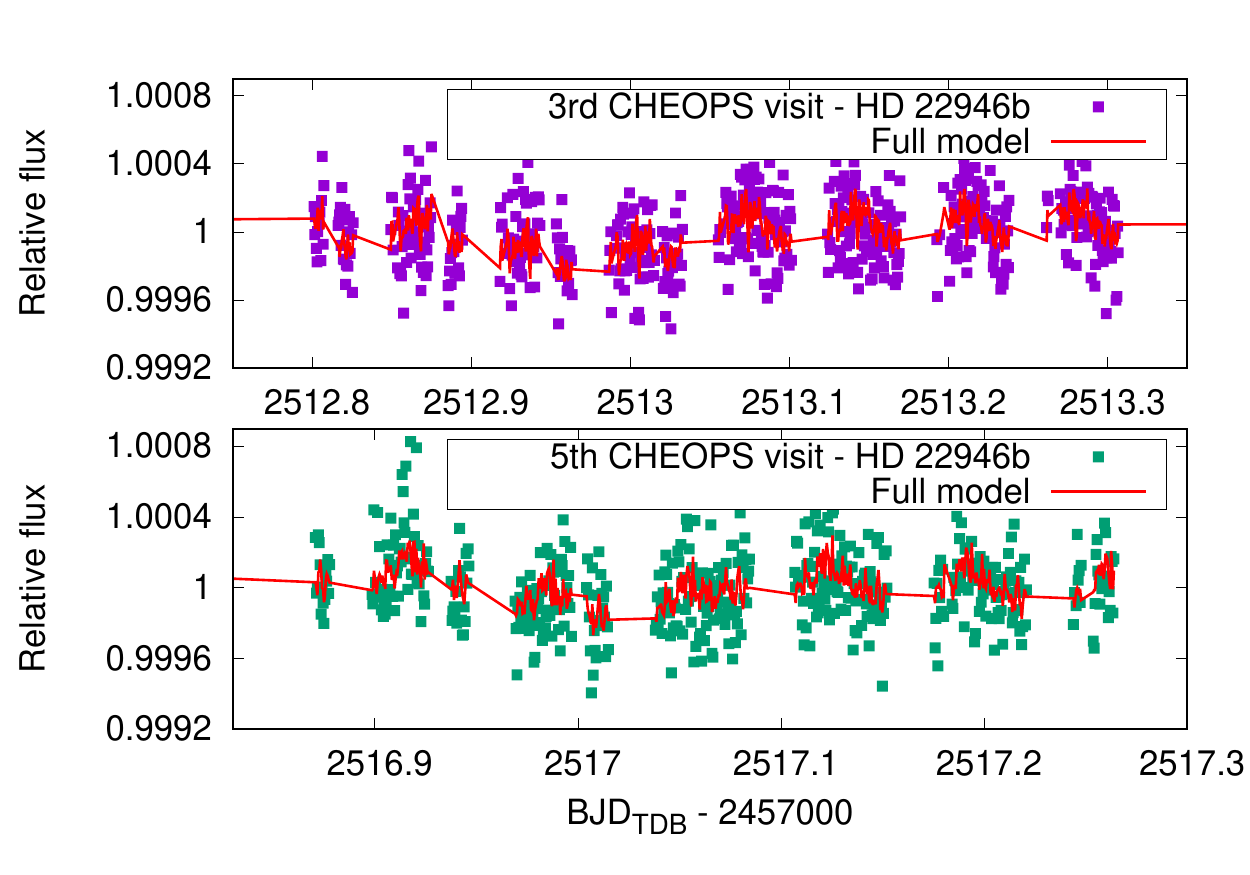}
\includegraphics[width=\columnwidth]{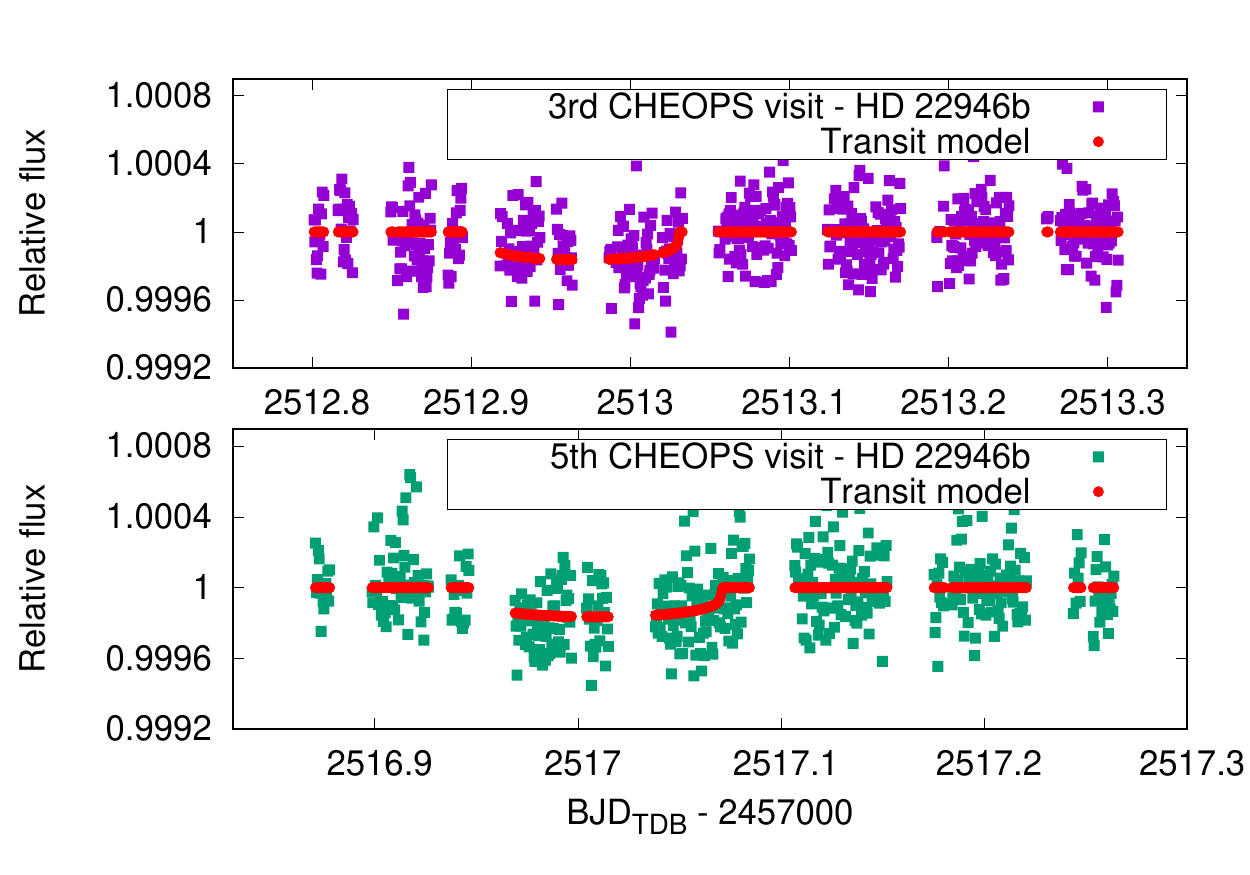}}
\centerline{
\includegraphics[width=\columnwidth]{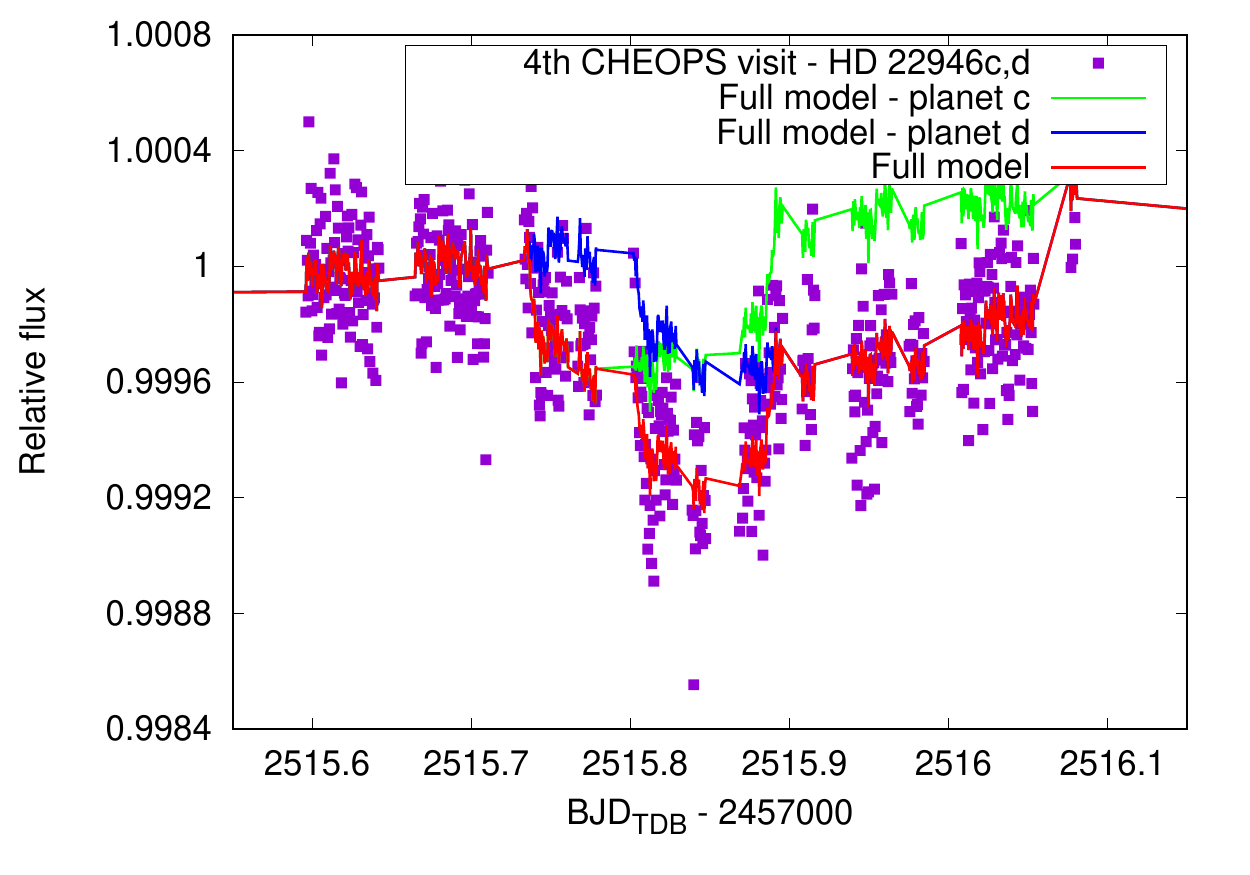}
\includegraphics[width=\columnwidth]{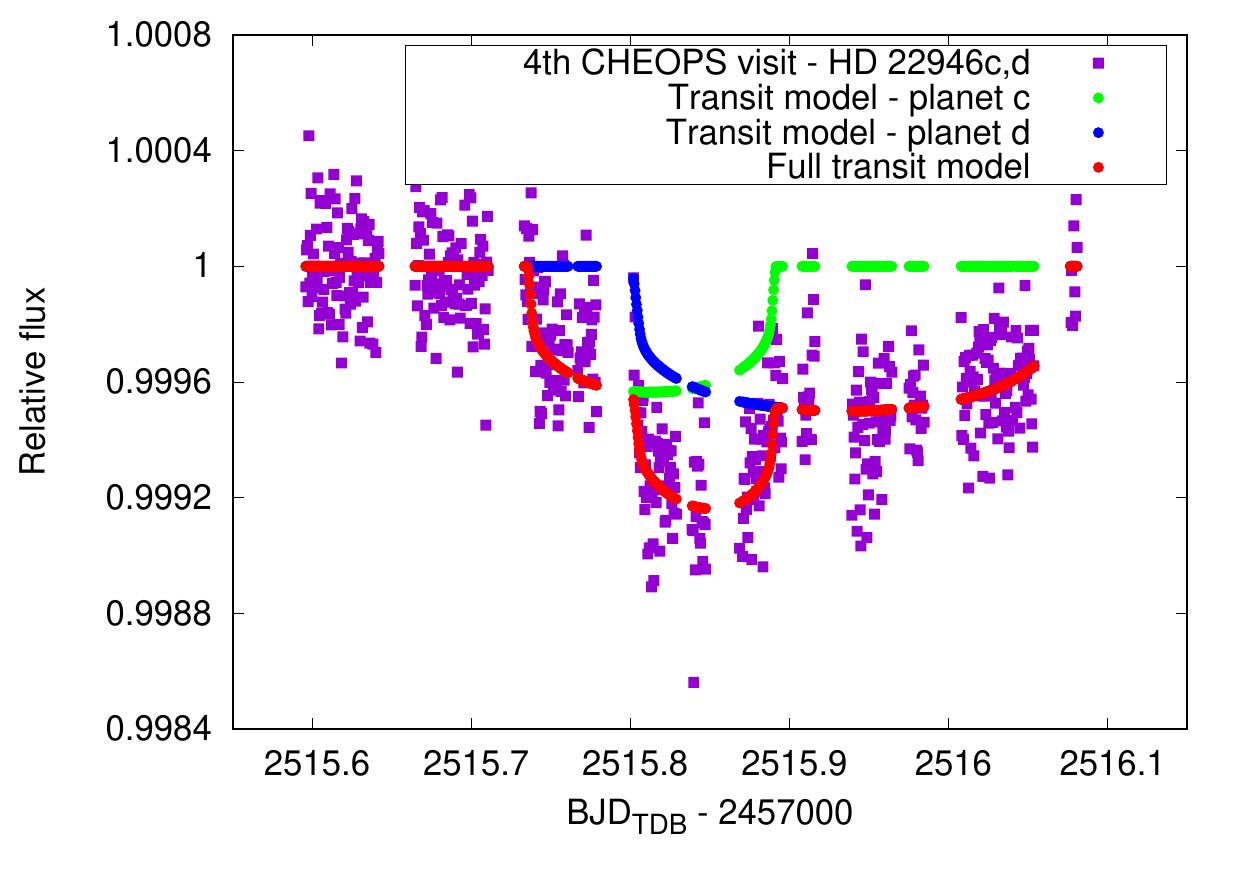}}
\caption{Individual CHEOPS observations of the transiting planets HD\,22946b, HD\,22946c, and HD\,22946d. The observed light curves are overplotted with the best-fitting model. This model was derived based on the entire CHEOPS and TESS photometric dataset and the RV observations from ESPRESSO via joint analysis of the data. The left-hand panels show the non-detrended data overplotted with the full model, while the right-hand panels show the detrended data overplotted with the transit model. In the case of the fourth CHEOPS visit, as the multiple transit feature, the individual transit models of planets c and d are also shown in addition to the summed model.}
\label{cheopsmodel} 
\end{figure*}

\begin{figure}
\includegraphics[width=\columnwidth]{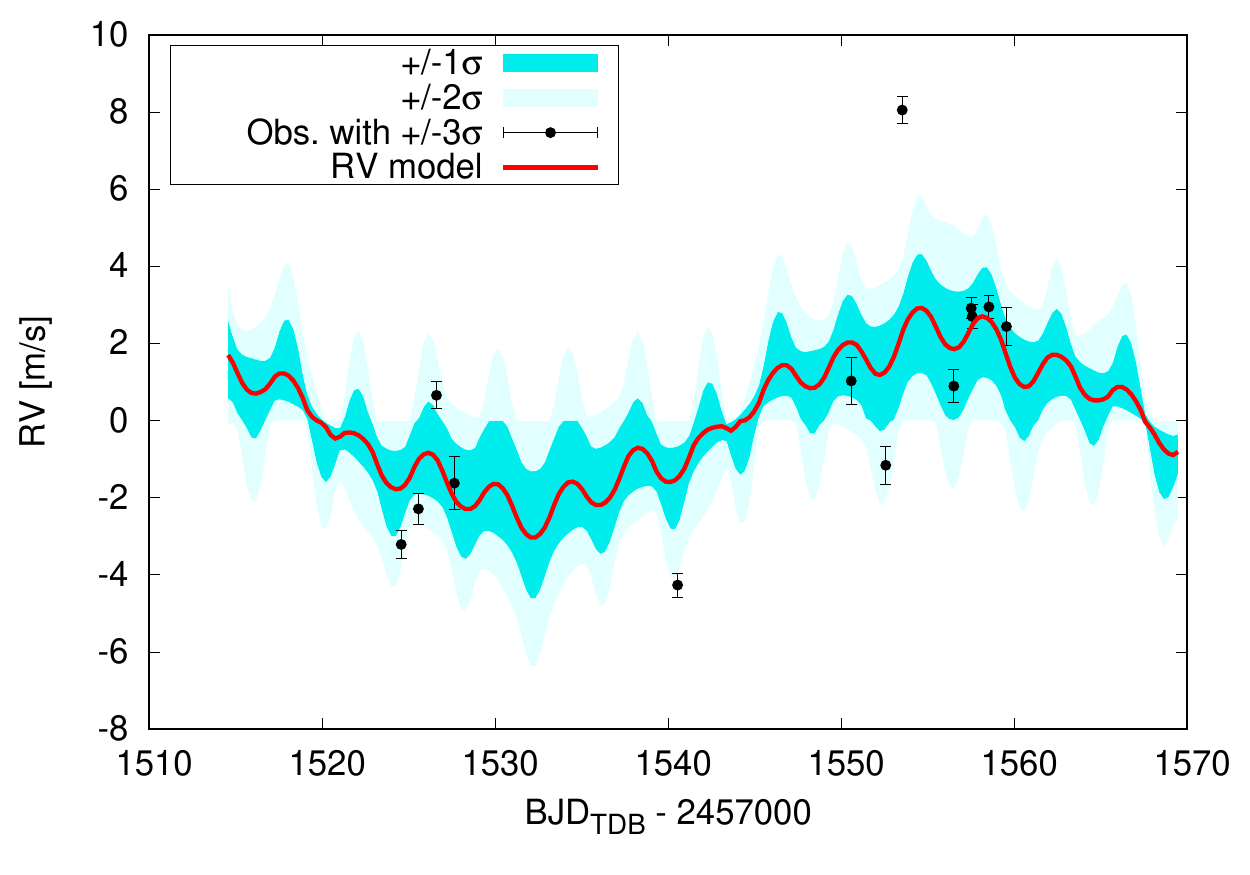}
\caption{RV observations taken at ESPRESSO fitted with a spectroscopic orbit (red line). The $1\sigma$ and $2\sigma$ uncertainties of the model are plotted as coloured areas. The uncertainties of the individual RV data points correspond to a $3\sigma$ interval.}
\label{rvdataplot} 
\end{figure}

The best-fitting and derived parameters from the combined model are listed in Table \ref{finalparams}, and the model posteriors of the host star are summarised in Appendix \ref{stelpost}. The fitted TESS light curves from sector numbers 3, 4, 30, and 31 are depicted in the panels of Figs. \ref{tessdataplot1} and \ref{tessdataplot2}. The CHEOPS individual observations overplotted with the best-fitting models are shown in the panels of Fig. \ref{cheopsmodel}. The RV observations fitted with a spectroscopic orbit are depicted in Fig. \ref{rvdataplot}.

Here, we present new ephemerides of the planetary orbits, which we calculated based on the combined model. Thanks to the combined TESS and CHEOPS observations, we were able to improve the reference mid-transit times and the orbital periods of the planets compared to the discovery values. C22 derived the orbital period parameter values of $P_\mathrm{orb,b} = 4.040301^{+0.000023}_{-0.000042}$ d and $P_\mathrm{orb,c} = 9.573096^{+0.000026}_{-0.000023}$ d, and expected an orbital period of $P_\mathrm{orb} = 46 \pm 4$ d 
for planet d, which was estimated based on the transit duration and depth along with stellar mass and radius through Kepler's third law, assuming circular orbits. We confirmed this prediction, finding an orbital period for planet d of $P_\mathrm{orb} = 47.42489 \pm 0.00011$ d. The improved ratios of the orbital periods are $P_\mathrm{orb,c}/P_\mathrm{orb,b} = 2.37$ and $P_\mathrm{orb,d}/P_\mathrm{orb,c} = 4.95$. Based on the \textit{Kepler} database, the adjacent planet pairs in multiple systems show a broad overall peak between period ratios of 1.5 and 2, followed by a declining tail to larger period ratios. In addition, there appears to be a sizeable peak just interior to the period ratio 5 \citep{Steffen1}; therefore, we can say that the period ratios in HD\,22946 fall into statistics and the seemingly large orbital gap between planets c and d is not anomalous.             

In the combined model, we determined the impact parameter $b$, which is the projected relative distance of the planet from the stellar disk centre during the transit midpoint in units of $\mathrm{R}_\mathrm{s}$. Converting these parameter values to the orbit inclination angle values we can obtain $i = 88.90^{+0.16}_{-0.05}$ deg, $i = 88.52^{+0.08}_{-0.07}$ deg, and $i = 89.54^{+0.02}_{-0.03}$ deg for planets b, c, and d, respectively. For comparison, we note that the corresponding discovery values are $i_\mathrm{b} = 88.3^{+1.1}_{-1.2}$ deg and $i_\mathrm{c} = 88.57^{+0.86}_{-0.53}$ deg. The inclination angle of planet d was not determined by C22. According to the improved parameter values, it seems that only the orbits of planets b and c are well aligned. Planet d is probably not in the same plane as planets b and c. 

Based on the combined TESS and CHEOPS photometry observations, we redetermined the radii of the planets, which are $1.362 \pm 0.040~\mathrm{R_\oplus}$, $2.328 \pm 0.039~\mathrm{R_\oplus}$, and $2.607 \pm 0.060~\mathrm{R_\oplus}$ for planets b, c, and d, respectively. The CHEOPS observations are an added value, because compared to the corresponding parameter values presented in C22 ($R_\mathrm{p,b} = 1.72 \pm 0.10~\mathrm{R_\oplus}$, $R_\mathrm{p,c} = 2.74 \pm 0.14~\mathrm{R_\oplus}$, and $R_\mathrm{p,d} = 3.23 \pm 0.19~\mathrm{R_\oplus}$), there is a noticeable improvement in radius precision. Using TESS and CHEOPS photometry observations, the uncertainties on the planet radius parameter values were decreased by $\sim 50$\%, 68\%, and 61\% for planets b, c, and d, respectively. We also note that the parameter values from this work are in stark contrast to those derived by C22; these authors found significantly larger radii, that is, larger by  $\sim 21$\%, 15\%, and 19\% for planets b, c, and d, respectively. We believe this may be due to a misunderstanding of the \texttt{LimbDarkLightCurve} function in \texttt{exoplanet}. The function requires the planetary radius $R_\mathrm{p}$ in solar radii rather than the planet-to-star radius ratio $R_\mathrm{p}/R_\mathrm{s}$. When misused, the result is an inflation of $R_\mathrm{p}/R_\mathrm{s}$ and $R_\mathrm{p}$ values by a factor of $R_\mathrm{s}/R_{\oplus}$, which in this case is a factor of about 15\%--21\%. This mistake can be seen most clearly in C22, when comparing the models shown in Figure 5 with the implied depths in Table 4 (likely derived from the radius ratio), which are inflated by this factor. Such a mistake was also evident during the reanalysis of the BD+40 2790 (TOI-2076) system \citep{Osborn1}.

According to the radius valley at $\sim 1.5 - 2.0~\mathrm{R}_\oplus$, which separates super-Earths and sub-Neptunes \citep{Fulton1, vanEylen1, Martinez1, Ho1}, and based on the refined planet radii, we find that planet b is a super-Earth, and planets c and d are similar in size and are sub-Neptunes,  in agreement with C22. It is well known that small exoplanets have bimodal radius distribution separated by the radius valley. Potential explanations focus on atmospheric-escape-driven mechanisms, such as photo-evaporation; see for example \citet{Owen1}. The models showed that those planets that have radius below $1.5~\mathrm{R}_\oplus$ were planets that initially had hydrogen/helium atmospheres, but ultimately lost them due to atmospheric escape, while those just above $2.0~\mathrm{R}_\oplus$ had hydrogen/helium atmosphere masses of $\sim 1$\% of the core mass. Having HD\,22946 planets on either side of the valley means that planet b could be a photo-evaporated version of planets c and d. Recently, \citet{luque2022density} presented a brand new approach, arguing that the density of planets might provide more information than planet radii alone and proposing that a density gap separates rocky from water-rich planets. For M dwarf systems, these authors found that rocky planets form within the ice line while water worlds formed beyond the ice line and migrated inwards. Given that theoretical models predict similar results  for stars of other types, this scenario could also be possible in the case of the planets orbiting HD\,22946.  

Due to the low number of RVs, here we present only the $3\sigma$ upper limits for the planet masses in agreement with the discoverers. C22 obtained the $3\sigma$ upper mass limits of about $11~\mathrm{M}_\oplus$, $14.5~\mathrm{M}_\oplus$, and $24.5~\mathrm{M}_\oplus$ for planets b, c, and d, respectively, from the same spectroscopic observations. The $3\sigma$ upper limits for the planet masses from this work are $M_\mathrm{p,b} = 13.71~\mathrm{M}_\oplus$, $M_\mathrm{p,c} = 9.72~\mathrm{M}_\oplus$, and $M_\mathrm{p,d} = 26.57~\mathrm{M}_\oplus$. Similarly to the discoverers, we obtained very different upper mass limits for planets c and d, although they have similar planet radii, which could be due to a somewhat different internal structure of these planets. Applying the relations of \citet{Chen1} and \citet{Otegi1}, we also re-estimated the planet masses, which were previously forecasted by the discoverers as $6.29 \pm 1.30~\mathrm{M}_\oplus$, $7.96 \pm 0.69~\mathrm{M}_\oplus$, and $10.53 \pm 1.05~\mathrm{M}_\oplus$ for planets b, c, and d, respectively. The improved parameter values are presented in Table \ref{finalparams}. Furthermore, taking into account the estimated planet masses calculated based on the relations of \citet{Otegi1}, we predicted the number of additional RV measurements required to achieve a $3\sigma$ detection on each mass using the \texttt{Radial Velocity Follow-up Calculator}\footnote{See \url{http://maestria.astro.umontreal.ca/rvfc/}.} (\texttt{RVFC}; see \citet{Cloutier1}), and the \texttt{RV simulator} (Wilson et al. in preparation). Based on these simulations, we obtained that another 27, 24, and 48 ESPRESSO RVs are needed to measure the predicted masses of planets b, c, and d, respectively. The expected RV semi-amplitudes assuming the estimated planet masses are $K_\mathrm{b} = 1.10 \pm 0.12~\mathrm{m~s}^{-1}$, $K_\mathrm{c} = 2.08 \pm 0.10~\mathrm{m~s}^{-1}$, and $K_\mathrm{d} = 1.46 \pm 0.08~\mathrm{m~s}^{-1}$.            

C22 also probed the planets from the viewpoint of future atmospheric characterisation using the transmission spectroscopy metric (TSM); see Eq. 1 in \citet{Kempton1}. The authors obtained the TSM values of $65 \pm 10$, $89 \pm 16,$ and $67 \pm 14$ for planets b, c, and d, respectively. We revised these values based on the results from the present work. The improved TSM values (see Table \ref{finalparams}) do not satisfy the recommended value of TSM > 90 for planets with a radius of $1.5 < R_\mathrm{p} < 10~\mathrm{R_\oplus}$. On the other hand, given that this threshold is set very rigorously, in agreement with the discoverers, we can note that planet c could be a feasible target for transmission spectroscopy observations with future atmospheric characterisation missions, such as the planned \textit{Ariel} space observatory \citep{Tinetti1}.

Finally, we discuss the relevance of planet d among the known population of similar exoplanets. HD\,22946d represents a warm sub-Neptune. Based on the NASA Exoplanet Archive\footnote{See \url{https://exoplanetarchive.ipac.caltech.edu/index.html}.} \citep{Akeson1}, there are 5272 confirmed exoplanets up to 22 February 2023, but only 63 planets out of 5272 are sub-Neptune sized ($1.75 < R_\mathrm{p} < 3.5~\mathrm{R}_\oplus$) and transiting bright stars ($G \leq 10$ mag). Only 7 planets out of 63 have orbital periods longer than 30 days and only 4 planets out of 7 have an equilibrium temperature of below 550 K. Three planets have a lower insolation flux than planet d, namely TOI-2076d \citep{Osborn1}, HD\,28109d \citep{Dransfield1}, and HD\,191939 \citep{Badenas1}. HD\,22946d is therefore an interesting target for future follow-up observations. One of the questions to be answered in the near future is the composition and internal structure of sub-Neptune-type planets. Using CHEOPS observations, we determined the radius of planet d with high accuracy. Its true mass could be determined with another 48 ESPRESSO RV measurements according to the estimate we present above. A combination of mass and radius gives the overall density, which will be an important step forward towards understanding sub-Neptunes.               

\section{Conclusions}
\label{conc}

Based on the combined TESS and CHEOPS observations, we refined several parameters of the HD\,22946 planetary system. First of all, we improved the ephemerides of the planetary orbits in comparison with the discovery values. We can confirm that planets b and c have short orbital periods below 10 days, namely $4.040295 \pm 0.000015$ d and $9.573083 \pm 0.000014$ d, respectively. The third planet, HD\,22946d, has an orbital period of $47.42489 \pm 0.00011$ d, which we were able to derive based on additional CHEOPS observations. Furthermore, based on the combined TESS and CHEOPS observations, we derived precise radii for the planets, which are $1.362 \pm 0.040~\mathrm{R_\oplus}$, $2.328 \pm 0.039~\mathrm{R_\oplus}$, and $2.607 \pm 0.060~\mathrm{R_\oplus}$ for planets b, c, and d, respectively. On the one hand, we can confirm the conclusion of the discoverers that the planetary system consists of a super-Earth, and planets c and d are sub-Neptunes. On the other hand, we find the planet radii values to be in tension with the values presented in the discovery paper, which is very probably due to misuse of the software by the discoverers. The low number of ESPRESSO RV measurements allowed us to derive only the $3\sigma$ upper limits for the planet masses, which are $13.71~\mathrm{M}_\oplus$, $9.72~\mathrm{M}_\oplus$, and $26.57~\mathrm{M}_\oplus$ for planets b, c, and d, respectively. 

We also investigated the planets from the viewpoint of possible future follow-up observations. First of all, we can conclude that more RV observations are needed to improve the planet masses in this system. The applied spectroscopic observations allowed us to derive precise stellar parameters of the host star and to fit an initial spectroscopic orbit to the RV data, but there is ample room for improvement in this way. We estimated that another 48 ESPRESSO RVs are needed to measure the predicted masses of all planets in HD\,22946. Planet c could be a suitable target for future atmospheric characterisation via transmission spectroscopy. We can also conclude that planet d as a warm sub-Neptune is very interesting, because there are only a few similar confirmed exoplanets to date. Thanks to the synergy of TESS and CHEOPS missions, there is a growing sample of planets, such as HD\,22946d. Such objects are worth investigating in the near future, for example in order to investigate their composition and internal structure. Finally, we can mention that future photometric and/or spectroscopic observations could also be oriented to searching for further possible planets in this system.

\begin{acknowledgements}
We thank the anonymous reviewer for the helpful comments and suggestions. CHEOPS is an ESA mission in partnership with Switzerland with important contributions to the payload and the ground segment from Austria, Belgium, France, Germany, Hungary, Italy, Portugal, Spain, Sweden, and the United Kingdom. The CHEOPS Consortium would like to gratefully acknowledge the support received by all the agencies, offices, universities, and industries involved. Their flexibility and willingness to explore new approaches were essential to the success of this mission. This paper includes data collected with the TESS mission, obtained from the MAST data archive at the Space Telescope Science Institute (STScI). Funding for the TESS mission is provided by the NASA Explorer Program. STScI is operated by the Association of Universities for Research in Astronomy, Inc., under NASA contract NAS 5-26555. This research has made use of the Exoplanet Follow-up Observation Program (ExoFOP; DOI: 10.26134/ExoFOP5) website, which is operated by the California Institute of Technology, under contract with the National Aeronautics and Space Administration under the Exoplanet Exploration Program. This work has made use of data from the European Space Agency (ESA) mission {\it Gaia} (\url{https://www.cosmos.esa.int/gaia}), processed by the {\it Gaia} Data Processing and Analysis Consortium (DPAC, \url{https://www.cosmos.esa.int/web/gaia/dpac/consortium}). Funding for the DPAC has been provided by national institutions, in particular the institutions participating in the {\it Gaia} Multilateral Agreement. ZG acknowledges the support of the Hungarian National Research, Development and Innovation Office (NKFIH) grant K-125015, the PRODEX Experiment Agreement No. 4000137122 between the ELTE E\"{o}tv\"{o}s Lor\'and University and the European Space Agency (ESA-D/SCI-LE-2021-0025), the VEGA grant of the Slovak Academy of Sciences No. 2/0031/22, the Slovak Research and Development Agency contract No. APVV-20-0148, and the support of the city of Szombathely. GyMSz acknowledges the support of the Hungarian National Research, Development and Innovation Office (NKFIH) grant K-125015, a PRODEX Institute Agreement between the ELTE E\"{o}tv\"{o}s Lor\'and University and the European Space Agency (ESA-D/SCI-LE-2021-0025), the Lend\"{u}let LP2018-7/2021 grant of the Hungarian Academy of Science and the support of the city of Szombathely. ABr was supported by the SNSA. ACC acknowledges support from STFC consolidated grant numbers ST/R000824/1 and ST/V000861/1, and UKSA grant number ST/R003203/1. B.-O. D. acknowledges support from the Swiss State Secretariat for Education, Research and Innovation (SERI) under contract number MB22.00046. This project has received funding from the European Research Council (ERC) under the European Union’s Horizon 2020 research and innovation programme (project {\sc Four Aces}; grant agreement No 724427). It has also been carried out in the frame of the National Centre for Competence in Research PlanetS supported by the Swiss National Science Foundation (SNSF). DE acknowledges financial support from the Swiss National Science Foundation for project 200021\_200726. DG gratefully acknowledges financial support from the CRT foundation under Grant No. 2018.2323 "Gaseousor rocky? Unveiling the nature of small worlds". This work was also partially supported by a grant from the Simons Foundation (PI Queloz, grant number 327127). This work has been carried out within the framework of the NCCR PlanetS supported by the Swiss National Science Foundation under grants 51NF40\_182901 and 51NF40\_205606. IRI acknowledges support from the Spanish Ministry of Science and Innovation and the European Regional Development Fund through grant PGC2018-098153-B-C33, as well as the support of the Generalitat de Catalunya/CERCA programme. This work was granted access to the HPC resources of MesoPSL financed by the Region Ile de France and the project Equip@Meso (reference ANR-10-EQPX-29-01) of the programme Investissements d'Avenir supervised by the Agence Nationale pour la Recherche. KGI and MNG are the ESA CHEOPS Project Scientists and are responsible for the ESA CHEOPS Guest Observers Programme. They do not participate in, or contribute to, the definition of the Guaranteed Time Programme of the CHEOPS mission through which observations described in this paper have been taken, nor to any aspect of target selection for the programme. The Belgian participation to CHEOPS has been supported by the Belgian Federal Science Policy Office (BELSPO) in the framework of the PRODEX Program, and by the University of Liège through an ARC grant for Concerted Research Actions financed by the Wallonia-Brussels Federation; L.D. is an F.R.S.-FNRS Postdoctoral Researcher. LMS gratefully acknowledges financial support from the CRT foundation under Grant No. 2018.2323 ‘Gaseous or rocky? Unveiling the nature of small worlds’. This project was supported by the CNES. MF and CMP gratefully acknowledge the support of the Swedish National Space Agency (DNR 65/19, 174/18). M.G. is an F.R.S.-FNRS Senior Research Associate. ML acknowledges support of the Swiss National Science Foundation under grant number PCEFP2\_194576. NAW acknowledges UKSA grant ST/R004838/1. This work was supported by FCT - Fundação para a Ciência e a Tecnologia through national funds and by FEDER through COMPETE2020 - Programa Operacional Competitividade e Internacionalizacão by these grants: UID/FIS/04434/2019, UIDB/04434/2020, UIDP/04434/2020, PTDC/FIS-AST/32113/2017 \& POCI-01-0145-FEDER- 032113, PTDC/FIS-AST/28953/2017 \& POCI-01-0145-FEDER-028953, PTDC/FIS-AST/28987/2017 \& POCI-01-0145-FEDER-028987, O.D.S.D. is supported in the form of work contract (DL 57/2016/CP1364/CT0004) funded by national funds through FCT. PM acknowledges support from STFC research grant number ST/M001040/1. We acknowledge support from the Spanish Ministry of Science and Innovation and the European Regional Development Fund through grants ESP2016-80435-C2-1-R, ESP2016-80435-C2-2-R, PGC2018-098153-B-C33, PGC2018-098153-B-C31, ESP2017-87676-C5-1-R, MDM-2017-0737 Unidad de Excelencia Maria de Maeztu-Centro de Astrobiologí­a (INTA-CSIC), as well as the support of the Generalitat de Catalunya/CERCA programme. The MOC activities have been supported by the ESA contract No. 4000124370. SH gratefully acknowledges CNES funding through the grant 837319. S.C.C.B. acknowledges support from FCT through FCT contracts nr. IF/01312/2014/CP1215/CT0004. S.G.S. acknowledge support from FCT through FCT contract nr. CEECIND/00826/2018 and POPH/FSE (EC). ACC and TW acknowledge support from STFC consolidated grant numbers ST/R000824/1 and ST/V000861/1, and UKSA grant number ST/R003203/1. V.V.G. is an F.R.S-FNRS Research Associate. XB, SC, DG, MF and JL acknowledge their role as ESA-appointed CHEOPS science team members. YA and MJH acknowledge the support of the Swiss National Fund under grant 200020\_172746. LBo, VNa, IPa, GPi, RRa and GSc acknowledge support from CHEOPS ASI-INAF agreement n. 2019-29-HH.0. NCS acknowledges support from the European Research Council through the grant agreement 101052347 (FIERCE). This work was supported by FCT - Fundação para a Ciência e a Tecnologia through national funds and by FEDER through COMPETE2020 - Programa Operacional Competitividade e Internacionalização by these grants: UIDB/04434/2020; UIDP/04434/2020. AT thanks the Science and Technology Facilities Council (STFC) for a PhD studentship. P.E.C. is funded by the Austrian Science Fund (FWF) Erwin Schroedinger Fellowship, program J4595-N.
\end{acknowledgements}

\bibliographystyle{aa} 
\bibliography{Yourfile} 

\appendix

\section{Additional tables}
\label{hyperanddecorrparams}

\begin{table*}
\footnotesize
\centering
\caption{Best-fitting GP hyperparameters and detrending parameters.}
\label{hyppar}
\begin{tabular}{lll}
\hline
\hline
Parameter [unit]                                                                        & Description                                                                                                             & $\mathrm{Value}^{+1 \sigma}_{-1 \sigma}$\\   
\hline
\hline
$\log \sigma_\mathrm{CHEOPS}$ [ppt]                                     & log of scatter in the CHEOPS data                                                                        & $-2.384^{+0.057}_{-0.061}$\\
$\log \sigma_\mathrm{TESS}$ [ppt]                                               & log of scatter in the TESS data                                                                                  & $-2.845^{+0.024}_{-0.023}$\\
$\varphi_{\omega_0,\mathrm{CHEOPS}}$ [1/deg]                            & typical frequency of variation in the shared CHEOPS roll-angle GP                                 & $1.70^{+0.19}_{-0.16}$\\
$\omega_\mathrm{0,TESS}$ [1/d]                                          & frequency for the TESS \texttt{SHOTerm} GP kernel                                                        & $1.87^{+0.10}_{-0.09}$\\
$\varphi_\mathrm{S_0,CHEOPS}$                                           & scaled power in the shared CHEOPS roll-angle GP                                                   & $0.0062^{+0.0010}_{-0.0009}$\\
$S_\mathrm{0,TESS}$                                                             & scaled power in the TESS \texttt{SHOTerm} GP kernel                                              & $0.0066^{+0.0010}_{-0.0008}$\\
$\varphi_\mathrm{power,CHEOPS}$                                         & power in the shared CHEOPS roll-angle GP                                                                  & $0.052^{+0.020}_{-0.013}$\\
$\varphi_\mathrm{lengthscale,CHEOPS}$ [deg]                     & typical lengthscale of variation in the shared CHEOPS rollangle GP                               & $3.68^{+0.38}_{-0.37}$\\
$df/d(\mathrm{time})_\mathrm{1st,CHEOPS}$                       & linear detrending parameter for time trend vs. flux -- 1st CHEOPS visit                         & $0.075 \pm 0.011$\\
$df/d(\mathrm{bg})_\mathrm{1st,CHEOPS}$                         & linear detrending parameter for background vs. flux -- 1st CHEOPS visit                 & $-0.026 \pm 0.036$\\
$d^2f/d(\mathrm{bg})^2_\mathrm{1st,CHEOPS}$                     & quadratic detrending parameter for background vs. flux -- 1st CHEOPS visit        & $0.0278 \pm 0.0025$\\
$df/d(\mathrm{centroid~x})_\mathrm{1st,CHEOPS}$                 & linear detrending parameter for x centroid vs. flux -- 1st CHEOPS visit                         & $0.0386^{+0.0080}_{-0.0078}$\\
$df/d(\mathrm{centroid~y})_\mathrm{1st,CHEOPS}$                 & linear detrending parameter for y centroid vs. flux -- 1st CHEOPS visit                         & $-0.0458^{+0.0083}_{-0.0084}$\\
$df/d(\mathrm{centroid~x})_\mathrm{2nd,CHEOPS}$         & linear detrending parameter for x centroid vs. flux -- 2nd CHEOPS visit                   & $0.202^{+0.044}_{-0.045}$\\
$d^2f/d(\mathrm{centroid~x})^2_\mathrm{2nd,CHEOPS}$     & quadratic detrending parameter for x centroid vs. flux -- 2nd CHEOPS visit           & $-0.0304 \pm 0.0083$\\
$df/d(\mathrm{time})_\mathrm{3rd,CHEOPS}$                       & linear detrending parameter for time trend vs. flux -- 3rd CHEOPS visit                         & $0.015 \pm 0.010$\\
$df/d(\mathrm{centroid~y})_\mathrm{3rd,CHEOPS}$                 & linear detrending parameter for y centroid vs. flux -- 3rd CHEOPS visit                         & $-0.0357^{+0.0074}_{-0.0075}$\\
$df/d(\mathrm{time})_\mathrm{4th,CHEOPS}$                       & linear detrending parameter for time trend vs. flux -- 4th CHEOPS visit                         & $0.103 \pm 0.013$\\
$df/d(\mathrm{centroid~y})_\mathrm{4th,CHEOPS}$                 & linear detrending parameter for y centroid vs. flux -- 4th CHEOPS visit                         & $-0.061 \pm 0.010$\\
$d^2f/d(\mathrm{centroid~y})^2_\mathrm{4th,CHEOPS}$     & quadratic detrending parameter for y centroid vs. flux -- 4th CHEOPS visit           & $0.0086 \pm 0.0019$\\
$x_\mathrm{mean,1st,CHEOPS}$ [ppt]                                      & the average offset of the median-normalised flux -- 1st CHEOPS visit                             & $0.007 \pm 0.011$\\
$x_\mathrm{mean,2nd,CHEOPS}$ [ppt]                              & the average offset of the median-normalised flux -- 2nd CHEOPS visit                & $0.066 \pm 0.011$\\
$x_\mathrm{mean,3rd,CHEOPS}$ [ppt]                                      & the average offset of the median-normalised flux -- 3rd CHEOPS visit                     & $0.039 \pm 0.010$\\
$x_\mathrm{mean,4th,CHEOPS}$ [ppt]                                      & the average offset of the median-normalised flux -- 4th CHEOPS visit                     & $0.314 \pm 0.014$\\
$x_\mathrm{mean,5th,CHEOPS}$ [ppt]                                      & the average offset of the median-normalised flux -- 5th CHEOPS visit                     & $0.061 \pm 0.011$\\
$x_\mathrm{mean,TESS}$ [ppt]                                            & the average offset of the median-normalised flux in TESS data                                    & $0.0179^{+0.0085}_{-0.0084}$\\
\hline
\hline
\end{tabular}
\tablefoot{Based on the joint fit of the TESS and CHEOPS photometric data, and RV observations. The table shows the best-fitting value of the given parameter and its $\pm 1\sigma$ uncertainty. The fitted values correspond to quantile 0.50 (median) and the uncertainties to quantils $\pm 0.341$ in the parameter distributions obtained from the samples.}
\end{table*}

\begin{table}
\centering
\caption{As in Table \ref{hyppar}, but for model posteriors of the host HD\,22946.}
\label{stelpost}
\begin{tabular}{lll}
\hline
\hline
Parameter [unit] & Description & $\mathrm{Value}^{+1 \sigma}_{-1 \sigma}$\\
\hline
\hline
$T_\mathrm{eff}$ [K]                                            & effective temperature             & $6167^{+64}_{-63}$\\
$\log g$ [cgs]                                                  & log of surface gravity                 & $4.358 \pm 0.016$\\
$u_\mathrm{1,TESS}$                                     & quadratic LD coefficient         & $0.196 \pm 0.077$\\
$u_\mathrm{2,TESS}$                                     & quadratic LD coefficient         & $0.244 \pm 0.086$\\
$u_\mathrm{1,CHEOPS}$                                   & quadratic LD coefficient         & $0.346^{+0.081}_{-0.079}$\\
$u_\mathrm{2,CHEOPS}$                                   & quadratic LD coefficient         & $0.291^{+0.089}_{-0.090}$\\
$R_\mathrm{s}$ [$\mathrm{R}_\odot$]                         & stellar radius                         & $1.1161^{+0.0088}_{-0.0089}$\\
$M_\mathrm{s}$ [$\mathrm{M}_\odot$]                         & stellar mass                         & $1.105 \pm 0.038$\\
$\gamma$ [$\mathrm{m~s}^{-1}$]                              & system RV                          & $16~854.48^{+0.82}_{-0.76}$\\ 
\hline
\hline
\end{tabular}
\end{table}

\end{document}